\documentclass[aps,pre,twocolumn,groupedaddress,showpacs,floatfix,superscriptaddress,longbibliography]{revtex4-2}
\usepackage{amsmath}
\usepackage{amssymb}
\usepackage{physics}
\usepackage{graphicx}
\usepackage{tabularx}
\usepackage[dvipsnames]{xcolor}
\usepackage[normalem]{ulem}
\usepackage{soul} 
\usepackage{ragged2e}
\usepackage{hyperref}
\usepackage{booktabs}
\usepackage{orcidlink}

\usepackage[caption=false]{subfig}
\MakeRobust{\subref}

\emergencystretch=\maxdimen
\hypersetup{
 colorlinks=true,
 linkcolor=blue,
 anchorcolor = blue,
 citecolor = blue,
 filecolor = blue,
 urlcolor = blue
}

\newcommand{\oist}{Okinawa Institute of Science and Technology Graduate University, Onna-son, Okinawa 904-0495, Japan}

\footnotetext[3]{Both authors contributed equally to this work.}

\begin{document}
\title{Fractal Path Strategies for Efficient 2D DMRG Simulations} 

\author{Oliver R. Bellwood \,\orcidlink{0000-0002-0691-4175} \ddag}
\email{oliver-bellwood@oist.jp}
\affiliation{\oist}

\author{Heitor P. Casagrande \,\orcidlink{0000-0003-0247-339X} \ddag}
\email{heitor-peres@oist.jp}
\affiliation{\oist}

\author{William J. Munro \,\orcidlink{0000-0003-1835-2250}}
\affiliation{\oist}

\begin{abstract} 
Numerical simulations of quantum magnetism in two spatial dimensions are often constrained by the area law of entanglement entropy, which heavily limits the accessible system sizes in tensor network methods. In this work, we investigate how the choice of mapping from a two-dimensional lattice to a one-dimensional path affects the accuracy of the two-dimensional Density Matrix Renormalization Group algorithm. We systematically evaluate all mappings corresponding to a subset of the Hamiltonian paths of the $N \cross N$ grid graphs up to $N=8$ and demonstrate that the fractal space-filling curves generally lead to faster convergence in ground state searches compared to the commonly used ``snake" path. To explain this performance gain, we analyze various locality metrics and propose a scalable method for constructing high-performing paths on larger lattices by tiling smaller optimal paths. Our results show that such paths consistently improve simulation convergence, with the advantage increasing with system size.
\end{abstract}
  
\maketitle

\setcounter{figure}{0}


\section{Introduction}\label{Intro}

The study of quantum magnetism in one-dimension (1D) benefits from some of the most powerful analytic tools available in many-body physics. Techniques such as the Bethe ansatz \cite{bethe1931theorie}, Tomonaga-Luttinger liquid theory \cite{FDMHaldane_1981}, and \hbox{$1+1$} dimensional conformal field theory \cite{BELAVIN1984333} owe much of their success to the geometric constraint of a single spatial dimension \cite{giamarchi}. In contrast, the study of quantum magnetism in two-dimensions (2D) remains a formidable challenge \cite{LowDMilestones,Schmidt2017,DagottoComplexity,Balents2010SpinLiquids}. Spin-wave theoretic approaches can achieve reasonable accuracy in 2D, where quantum fluctuations act only as small corrections to an ordered ground state \cite{Oguchi,Hamer}. However, these techniques tend to break down when fluctuations become significant, which is often the case in both 1D and 2D systems \cite{LimitsLSWT,auerbach}. Moreover, topological properties lie at the heart of many exotic phases in 2D \cite{BernevigTopology}. While quasiparticles in 1D are prohibited from exhibiting braiding statistics, such phenomena are fundamental to the physics of 2D quantum systems. As a result, quantum magnetism in 2D occupies a ``Goldilocks Zone" of theoretical difficulty: quantum fluctuations and topological effects are both substantial, yet the powerful analytical tools of 1D do not readily generalize.

In light of these analytical challenges, computational methods provide a vital avenue for investigating quantum systems in 1D and 2D. Tensor network-based algorithms \cite{orus2014, montangero2018introduction, FelserMontangero2021, fishman2022itensor} such as the Density Matrix Renormalization Group (DMRG) \cite{White1993DMRG,schollwock2005,mcculloch2007,schollwock2011} have been extensively used for studying both closed \cite{PhysRevLett.133.190402, PhysRevLett.93.040502, Verstraete2023, PhysRevB.91.045115, wang2025symmetrizingconstraintsdensity, PhysRevLett.128.230401} and open \cite{PhysRevLett.106.220601, PhysRevLett.114.220601, PhysRevLett.115.080604, Casagrande2021, PhysRevA.92.022116} quantum systems in 1D. In 2D, Methods based on Projected Entangled Pair States (PEPS) \cite{PEPS} and the Multiscale Entanglement Renormalization Ansatz (MERA) \cite{MERA2} have achieve accuracies comparable to those well-established techniques such as quantum Monte Carlo (QMC) \cite{Sandvik1997QMC} and exact diagonalization (ED) \cite{Schulz1996ED}. Importantly, tensor network methods are not subject to the notorious ``sign problem" \cite{QMCSignProblem} that limits the applicability of QMC in many 2D systems. Despite being intrinsically one-dimensional, the DMRG algorithm has also demonstrated considerable success when extended to 2D \cite{DMRG2012twodimensions}. In the context of quantum magnetism, DMRG has been used to study frustrated and unfrustrated Heisenberg models on various lattice geometries \cite{White1994RVB,WhiteChernyshev2007}. It has also played a key role in the simulation and characterization of candidate quantum spin liquid phases in the $J_1-J_2$ square and triangular Heisenberg models \cite{Balents2012J1_J2_square, ZhuWhite2015_J1_J2Tri}, the Kagome lattice Heisenberg model \cite{Depenbrock2012Kagome}, and the Kitaev Honeycomb model \cite{Jiang2011KitaevHC}.

A key technical challenge in applying the DMRG algorithm to systems in more than one dimensions is the identification of a bijective mapping from the higher-dimensional Hamiltonian onto a one-dimensional model \cite{LiangPang1994}. This mapping enables the use of a powerful 1D ansatz by representing the two-dimensional many-body wavefunction as a Matrix Product State (MPS) \cite{perez2007MPS}, offering substantial flexibility and control in the calculation of physical properties. The strength of the MPS formalism lies in its ability to iteratively approximate the system's ground state energy within the DMRG framework. At each step, the algorithm performs a variational energy minimization, followed by truncation of the MPS's internal dimensions. The key computational parameter in this process is the bond dimension $D$, which determines the maximum matrix sizes in the truncated MPS \cite{schollwock2005, schollwock2011}. 

A fundamental consequence of mapping a two-dimensional system into one-dimension is that some interactions between nearest neighbors in the original lattice become long-ranged in the 1D representation, requiring larger bond dimension in the DMRG algorithm to maintain accuracy. This directly increases the computational cost \cite{LiangPang1994, Xiang2001DMRG2D, DMRG2012twodimensions}, for a given MPS and associated Matrix Product Operator (MPO). Moreover, one-dimensional systems whose bipartite entanglement scales according to the ``area law" \cite{Hastings_2007} can be efficiently simulated using tensor network ansatz \cite{PhysRevLett.100.030504}. In 2D, however, the area law implies that entanglement entropy scales with the linear size of the system, which imposes significant constraints on the efficiency of MPS representations \cite{Eisert2010AreaLaw}. As a result, 1D mappings that minimize the effective range of interactions are highly desirable, as they can lead to substantial improvements in both convergence speed and numerical precision.

In this work, we investigate the effect of different 2D to 1D mappings on the accuracy of numerical results from the DMRG algorithm. Specifically, we investigate the ground state properties of the spin-$1/2$ Heisenberg antiferromagnet on the square lattice \cite{Chakravarty1988SLHAF} using mappings corresponding to the Hamilton paths of the $N \cross N$ planar grid graph \cite{Jacobsen2007}. We compute the ground state energy, variance, and von-Neumann entanglement entropy for a subset of the Hamilton paths and determine the optimal paths that yield the greatest accuracy for a given amount of computational power for $2 \leq N \leq 8$.

We report that, for even $N$, the paths with the fastest convergence to the ground state are unambiguously and invariably fractal space-filling curves \cite{Sagan1994SF, FASS_paper, Bader2013SFCurves}. We suggest that this performance enhancement is likely related to the spatial locality-preserving nature of such curves \cite{JAGADISH1997, Cataldi2021hilbertcurvevs}, which allows for more efficient encoding of the large entanglement that is endemic to the 2D geometry of the problem. We build upon these findings by proposing and evaluating paths constructed by tiling the $N=5$ optimal path found in this work onto the finite order Hilbert curve \cite{FASS_paper} for the $N=10,20,$ and $40$ square lattices. We demonstrate that these paths greatly outperform the industry standard ``snake" path \cite{LiangPang1994, Li2019XTRG}, showing faster convergence to the ground state for the same amount of computational resources. 

\section{Model}

We consider the Hamiltonian of the $N \cross N$ square lattice spin-1/2 Heisenberg antiferromagnet (SLHAF) \cite{Chakravarty1988SLHAF}, given by  
\begin{equation} \label{SLHAF_Hamiltonian}
    \hat{H} = J \sum_{\langle i,j \rangle} \mathbf{S}_i \cdot \mathbf{S}_j,
\end{equation}
where the isotropic exchange interaction of strength $J > 0$ is only between nearest neighbors, denoted as $\langle ... \rangle$, $\mathbf{S}_i$ is the quantum spin operator on lattice site $i$ with $\hbar = 1$ henceforth, and we assume open boundary conditions (OBC). Currently, there is no exact analytic solution to this model \cite{RevModPhys_SLHAF}, in contrast to its direct one-dimensional counterpart, which is exactly solvable via the Bethe ansatz \cite{Takahashi_1999}. The single-band Hubbard Hamiltonian at half-filling is a commonly used effective Hamiltonian for the square-lattice cuprates \cite{Anderson1987CuO2}; in the undoped (Mott insulating) limit \cite{RevModPhys_SLHAF}, this effective Hamiltonian is equivalent to the SLHAF. Consequently, the SLHAF is a prototypical model of quantum magnetism which has seen substantial attention owing to its relation to the quasi-two-dimensional superconducting cuprates \cite{HighTC}. 

The ground state of the SLHAF shows conventional N\'eel long-ranged order \cite{RegerYoung1988QMC}, which is destroyed at $T>0$ in accordance with the Hohenberg-Mermin-Wagner Theorem \cite{Hohenberg, Mermin_Wagner}. In the absence of frustrating interactions or impurities the SLHAF falls within the `renormalized classical' regime, with an exponentially diverging correlation length for $T\rightarrow 0$ \cite{Chakravarty1988SLHAF}. In this regime the zero-point fluctuations around the N\'eel ordered ground state are small, which allows spin-wave theory to achieve appreciable accuracy in calculations of the ground state properties \cite{Hamer}. Additionally, the non-linear $\sigma$ model allows for a complete parametrization of the low-temperature behavior of the SLHAF. However, such parameters are inaccessible via theory and must be computed numerically \cite{Sandvik1997QMC}. This makes the SLHAF a good candidate for assessing the benefits of the informed choice of DMRG paths, as it possesses a rather well-behaved and conventional quantum many-body ground state that is only straightforwardly accessed via computational methods. 

\section{Hamilton Paths and Space Filling Curves}\label{Ham_paths_Curves}

A graph path between two vertices of a graph that visits every vertex only once is known as a Hamiltonian path; for the purposes of disambiguation with the Hamiltonian Operator we refer to these paths as Hamilton paths throughout this work. Determining whether an arbitrary graph hosts a Hamilton path (such graphs are termed `traceable') is known to be an NP-complete problem \cite{HPProblem_NPHard}. A Hamilton path on a lattice graph is a special case of a self-avoiding walk \cite{Guttman2001SAW}; these walks have been well studied in the context of protein folding, where they are referred to as compact polymers \cite{Flory1956, Oberdorf2006Compact}. This work concerns itself with the finite sized, planar, square grid graph. The exact enumeration of all Hamilton paths on the grid graph remains an open problem in combinatorial analysis \cite{Jacobsen2007}. Significant progress has been made via computational enumeration, where the number of Hamilton paths on the $N \cross N$ square lattice has been determined up to $N=17$ \cite{MGD1990, Jacobsen2007}. 

\begin{figure}[ht]
    \centering
    \subfloat[\label{Peano_2ndorder}]{%
        \includegraphics[width=0.46\columnwidth]{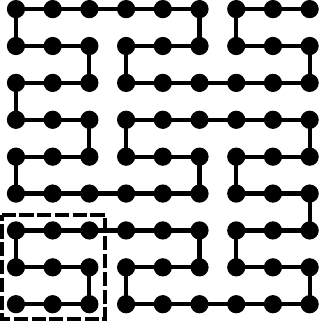}}
        \hfill
    \subfloat[\label{Hilbert_3rdorder}]{%
        \includegraphics[width=0.46\columnwidth]{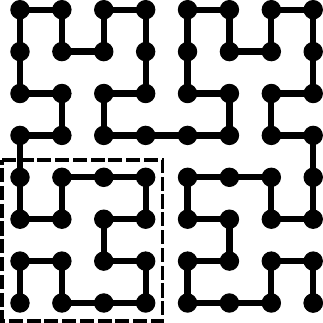}}
    
    \subfloat[\label{Hilb_motif}]{%
      \includegraphics[width=0.46\columnwidth]{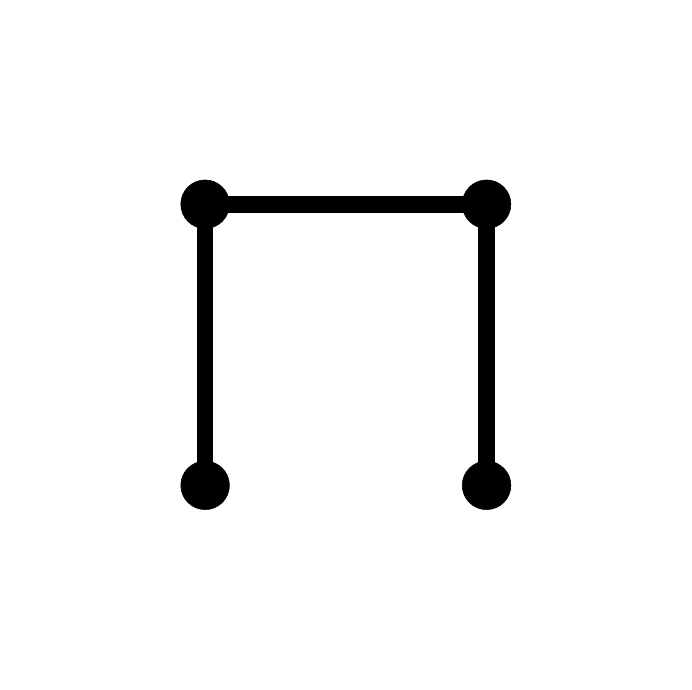}}
      \hfill
    \subfloat[\label{FASS_Hilb}]{%
      \includegraphics[width=0.44\columnwidth]{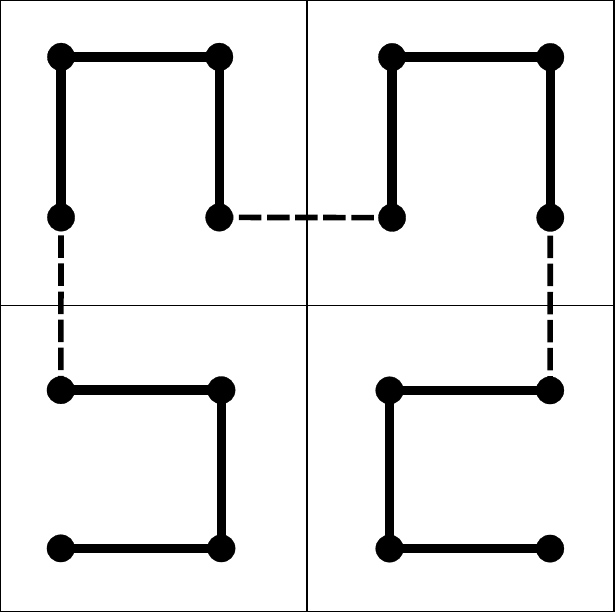}}
      
    \subfloat[\label{Meander_motif}]{%
      \includegraphics[width=0.46\columnwidth]{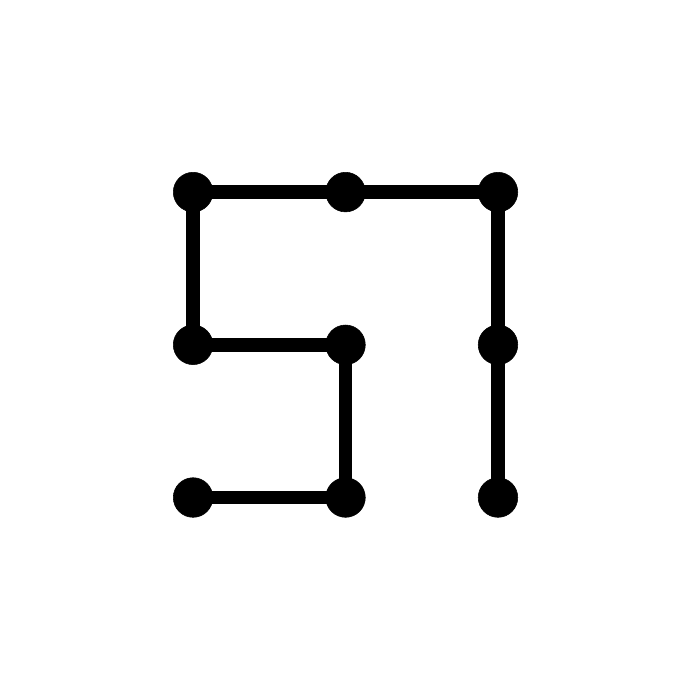}}
      \hfill
    \subfloat[\label{FASS_Meander}]{%
      \includegraphics[width=0.44\columnwidth]{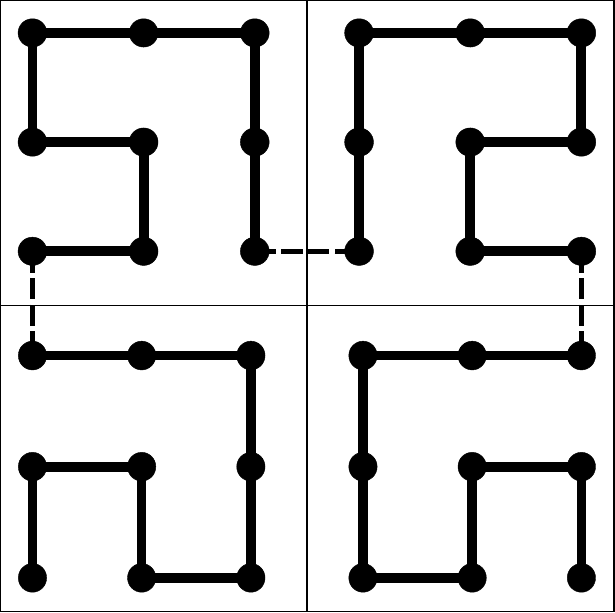}}
      
    \caption{\justifying Finite order, space filling curves of $N\cross N$ grid graphs are also Hamilton paths. \subref{Peano_2ndorder} The second order Peano curve is an $N=9$ Hamilton path and an ortho-tile. \subref{Hilbert_3rdorder} The third order Hilbert curve is an $N=8$ Hamilton path and a para-tile. The dashed squares denote the self-similar motif of the previous order that generates the curves. The first order \subref{Hilb_motif} and second order \subref{FASS_Hilb} Hilbert curves. A higher order curve is a tiling of the previous order curve, via rotations and translations. The endpoints must be connected, as shown with the dashed lines. The $N=3$ FASS para-tile motif \subref{Meander_motif} can be `Hilbert tiled' to generate an $N = 6$ Hamilton path \subref{FASS_Meander}; note that the lower quadrants differ from the motif in \subref{Meander_motif} via a reflection.}
    \label{PeanoHilbert}
\end{figure}

A Hamilton path through a lattice is equivalent to a bijective mapping between one- and two-dimensional space, where the ordering of lattice sites visited in the 1D path encodes a walk through the 2D graph. For an $N \cross N$ grid graph, there are $N^2!$ possible orderings of the vertices: only a minuscule number of these orderings correspond to the Hamilton paths, as the orderings must also obey the connectivity of the graph. The Hamilton paths therefore encompass all the unique ways that a two-dimensional lattice with nearest neighbor bonds can be mapped to a 1D path over said bonds. 

Finite order space filling curves of the $N \cross N$ grid graph are quintessential examples of Hamilton paths \cite{Sagan1994SF, Bader2013SFCurves}. Cantor showed that the unit interval, $[0,1]$, and the unit square, $[0,1]^2$, must have the same cardinality, which implies the possibility of mapping every point on $[0,1]^2$ to a (piece-wise) continuous one-dimensional object. Peano was the first to construct such a mapping in 1890, demonstrating the ability of curves, which are one-dimensional, to completely fill a two-dimensional space with finite area \cite{Peano1890}. In 1891 Hilbert offered a geometric construction of such objects, thereby allowing for a description of the entire class of space filling curves \cite{Hilbert_OG}. Their two eponymous curves, at second and third order respectively, are shown in Fig. \ref{PeanoHilbert}.  

The exploration of space-filling curves is not limited to the realm of mathematical theory. Many biological systems exhibit the self-similarity that is inherent to the space-filling curves \cite{Beauty_of_plants}. The Lindenmayer System, which was originally developed to model growth in filamentous organisms \cite{LINDENMAYER1968}, provides a compact grammar for describing all space-filling curves \cite{FASS_paper}. In computer science, higher dimensional data structures can be linearized using the mappings of the finite order space-filling curves; in doing so, the locality properties of the curves has been show to increase efficiency in load balancing and memory access \cite{Moon2001Analysis, Bader2013SFCurves}. 

All finite-order space-filling curves fall within the class of curves known as the approximately space-filling, self-avoiding, simple and self-similar curves, or `FASS' curves for short \cite{FASS_paper}. The FASS curves are abstractions of the geometric construction of the original space-filling curves as described by Hilbert, where one or more motifs may be recursively tiled onto the plane by connecting the endpoints of the motifs. For the $N \cross N$ grid graph, the number of unique motifs is equivalent to the number of symmetry-reduced Hamilton paths that start and end on corners. If the end points are on opposing corners these motifs are termed `ortho-tiles'; otherwise, the corners share a side of the graph and are termed `para-tiles' \cite{FASS_paper}. Unless stated we will always be referring to the para-tiles when discussing FASS motifs. The enumeration of the para-tile motifs up to $N=10$ is presented in Table \ref{paratiles_enumeration}. The tiling procedure as presented in Ref. \cite{FASS_paper} allows for the construction of larger sized FASS curves from smaller motifs. A tiling corresponding to the first order Hilbert curve (which we shall call a `Hilbert tiling') of the $N=3$ Peano-meander motif is shown in Fig. \ref{Meander_motif}. We will later see that this curve is exactly the optimal DMRG path for the $N=6$ SLHAF. This implies a procedure for generating high-performing DMRG paths via FASS tiling of smaller optimal paths. 

\begin{table}[htb]
    \centering
    \begin{tabularx}{\columnwidth} { 
    >{\raggedright\arraybackslash}X  
    >{\raggedleft\arraybackslash}X 
    >{\raggedleft\arraybackslash}X }
    \toprule
    N              & FASS para-tile motifs & Symmetry-reduced, corner starting\\
    \midrule
    1              & 1 & 1\\
    2              & 1 & 1\\
    3              & 1 & 3\\
    4              & 5 & 23\\
    5              & 43 & 347\\
    6              & 897 & 10199\\
    7              & 44209 & 683277\\
    8              & 4467927 & 85612967\\
    9              & 1043906917 & 25777385143\\
    10              & 506673590576 & 14396323278040\\
    \bottomrule
    \end{tabularx}
  \caption{\justifying The number of para-tile motifs and symmetry-reduced, corner starting Hamilton paths on an $N \cross N$ square lattice, up to $N=10$ \cite{oeisA384173, oeisA363577}. }
  \label{paratiles_enumeration}
\end{table}

\section{Optimality and locality of Hamilton paths}

The Hamilton paths are a natural choice of mapping as they preserve the largest number of nearest-neighbor bonds. However, the DMRG algorithm does not require the site orderings to respect the connectivity of the original 2D lattice. In fact, some of the first discussions of optimal DMRG paths in 2D considered site orderings that traversed the square lattice diagonally \cite{Xiang2001DMRG2D, Li2019XTRG}. For computational convenience, we restrict our investigation to only the Hamilton path mappings. 

The optimality of a DMRG path is defined by its accuracy in computing a given physical quantity for a fixed amount of computational resource. For the present investigation, the computed physical quantities are the normalized ground state energy per site, $E_0/ J N^2$, with
\begin{align}
    E_0 = \mel{\psi_0}{\hat{H}}{\psi_{0}},
\end{align}
the variance in ground state energy, 
\begin{equation}
    \sigma_H = \bra{\psi_0} \hat{H}^2\ket{\psi_0} - {E_0}^2,
\end{equation}
and the total bond (von Neumann) entanglement entropy, $\sum_{b} S_E(b)$, where $\ket{\psi_0}$ is the ground state MPS, and 
\begin{equation}
    S_{E}(b) = \Tr(\rho_{b} \ln \rho_{b})
\end{equation} with reduced density matrix $\rho_{b}$ over bond $b$. 

For all quantities, a smaller value is generally indicative of greater accuracy at large values of bond dimension $D$. It was suggested in Ref. \cite{Li2019XTRG} that path optimality is linked to the number of lattice bonds that are intersected (cut) by a typical bipartition when contracting over a single MPO bond. The fewer lattice bonds that are intersected by the bipartition, the smaller the required MPO and bond dimension. This is closely related to a suggestion \cite{Cataldi2021hilbertcurvevs} that path optimality is associated with locality preservation, whereby a 2D to 1D mapping that minimizes long-ranged interactions would naturally require a smaller bond dimension \cite{LiangPang1994}. Both of these observations are deeply tied to the area law that governs the entanglement scaling for DMRG computation in 2D \cite{vidal2004, Hastings_2007, Wolf2008Arealaw, Eisert2010AreaLaw}. Unlike the 1D case, ground state entanglement scales with system size in 2D. As DMRG is a maximal entanglement algorithm, this greatly limits the size and system parameters that are accessible with current simulation capabilities. 

It follows from previous studies on path optimality \cite{Li2019XTRG, Cataldi2021hilbertcurvevs, LiangPang1994, Xiang2001DMRG2D} that mappings with a smaller number of long-distance interactions or mappings where the long-distance interactions are short relative to system size will result in greater performance. These aspects can be quantified by the calculation of \textit{locality metrics} which have been well studied in the context of space-filling curves and their applications in computer science \cite{Bially1969, PKK1992Peano, Dai2003Locality, MD1986Optimal,  Bader2013SFCurves}. Locality metrics are measures of proximity preservation, i.e., sites that are close together in 2D remain so under mapping to 1D \cite{Dai2003Locality}. The mapping of a site ($i$) on a 2D lattice to a site ($n$) in a 1D path is defined as 
\begin{equation} \label{Mapping}
    \mathcal{M}: i \in ([1,N], [1,N]) \longrightarrow n \in [1,N^2],
\end{equation}
where the 2D site $i = (x,y)$ with $x-$ and $y-$coordinates in $[1,N]$ is now a site in the 1D path with index $n$ in $[1,N^2]$. For a path $P$ on the square grid graph, we denote $P(n)$ as the cartesian position of the $n^{\text{th}}$ site along the path, and $d(P(n),P(m))$ and $d_1(P(n),P(m))$ as the Euclidean and Manhattan distances between sites $n$ and $m$ respectively. The Perez, Kamata, and Kawaguchi (PKK) metric \cite{PKK1992Peano} is defined as 
\begin{equation}
    L_{\text{PKK}}(P) = \sum_{n,m \in 1,...,N^2 |n<m} \frac{\abs{n-m}}{d(P(n),P(m))},
\end{equation}
which was used to study the locality and performance of space-filling curves when applied to image scanning. The Dai and Su (DS) metric \cite{Dai2003Locality} is defined as 
\begin{equation}
    L_{\text{DS}, \delta}(P) = \sum_{\substack{n,m \in 1,...,N^2 |n<m \\ \text{ and } \\d_1(P(n),P(m))=\delta}} \abs{n-m},
\end{equation}
with integer $\delta$, which was used to analytically investigate the locality of the Hilbert and z-order curves in two- and three-dimensions. Lastly, the Mitchison and Durbin (MD) metric \cite{MD1986Optimal} is defined as 
\begin{equation}
    L_{\text{MD},q}(P) = \sum_{\substack{n,m \in 1,...,N^2 |n<m \\ \text{ and } \\ d(P(n),P(m))=1}} \abs{n-m}^q,
\end{equation}
which was used to assess optimality of number orderings to reduce the edgesum of an $N \cross N$ array. All three of these metrics are concerned with an object known as the path `index difference', $\abs{n-m}$. In the DMRG algorithm, sites that are coupled with a large index difference are by definition those with long-ranged interactions. It follows that mappings with small locality metrics are expected to perform better.

\section{Path Informed DMRG} \label{PIDMRG}

\begin{figure}[t]
\centering
\subfloat[\label{snake_4x4}]{%
\includegraphics[width=0.3\columnwidth]{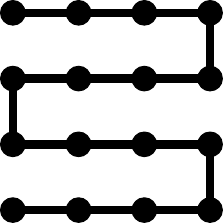}%
}\hfill
\subfloat[\label{spiral_4x4}]{%
  \includegraphics[width=0.3\columnwidth]{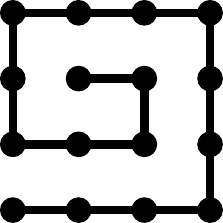}%
}\hfill
\subfloat[\label{Hilbert_4x4}]{%
  \includegraphics[width=0.3\columnwidth]{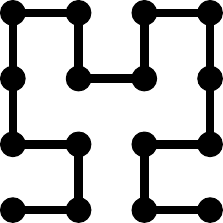}%
}
\caption{%
  \justifying   The snake path \subref{snake_4x4}, the spiral path \subref{spiral_4x4}, and the Hilbert curve path \subref{Hilbert_4x4} of the $4\cross4$ grid graph. }
    \label{3paths}
\end{figure}

As an illustrative example, let us consider the $4 \cross 4$ spin-1/2 SLHAF with OBC. In units of $J$, ED yields a ground state energy of $E_0^{\text{ED}}/J \approx -9.189$. There are 276 possible Hamilton paths through the $N = 4$ grid graph \cite{Jacobsen2007}. As the SLHAF Hamiltonian is spatially isotropic we may only concern ourselves with paths that are nonequivalent under reflection or rotations, as paths connected by a symmetry will yield the same result, since the mapping is functionally the same. There are 38 such symmetry-reduced Hamilton paths through the $N=4$ grid graph \cite{oeisA265914}. For the sake of computational convenience, we further restrict the paths to always have at least one starting point in a corner of the grid graph; in doing so, there are only 23 symmetry-reduced, corner starting Hamilton paths. For now, let us consider three well-defined paths of this subset: the snake path, the spiral path, and the Hilbert curve path, depicted in Fig. \ref{3paths}.

\begin{table}[ht]
  \centering
    \begin{tabular}{l*{6}{c}r}
    Path              & $L_{\text{PKK}} \quad$ & $L_{\text{DS}, 2}\quad$ & $L_{\text{MD},1}\quad$ & Total Bonds Cut\\
    \hline
    Snake              & \bf 316.1 & \bf 212 & \bf 60 & \bf 42  \\
    Spiral            & 358.0 & 288 & 88  & 99 \\
    Hilbert           & 321.9 & 228 & 64 & 49  \\
    \end{tabular}
  \caption{\justifying The locality metrics and number of bonds cut via bipartitions for the three identified paths on the $N=4$ square lattice. The lowest values for each metric are bolded, clearly indicating that the snake path is expected to have the best performance. }
  \label{3paths_localitytable}
\end{table}

\begin{figure}[!h]
\centering

\subfloat[\label{4x4_3paths}]{%
  \includegraphics[width=0.9\columnwidth,trim={0 1cm 0 0}]{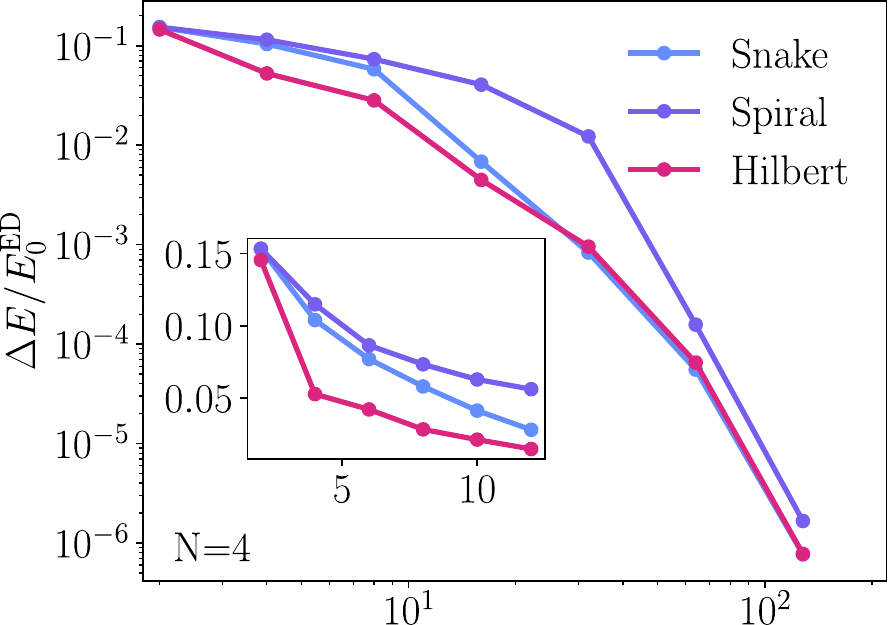}%
} \hfill
\subfloat[\label{4x4_SE_and_variance}]{%
  \includegraphics[width=0.87\columnwidth,trim={0 0 0.2cm 0}]{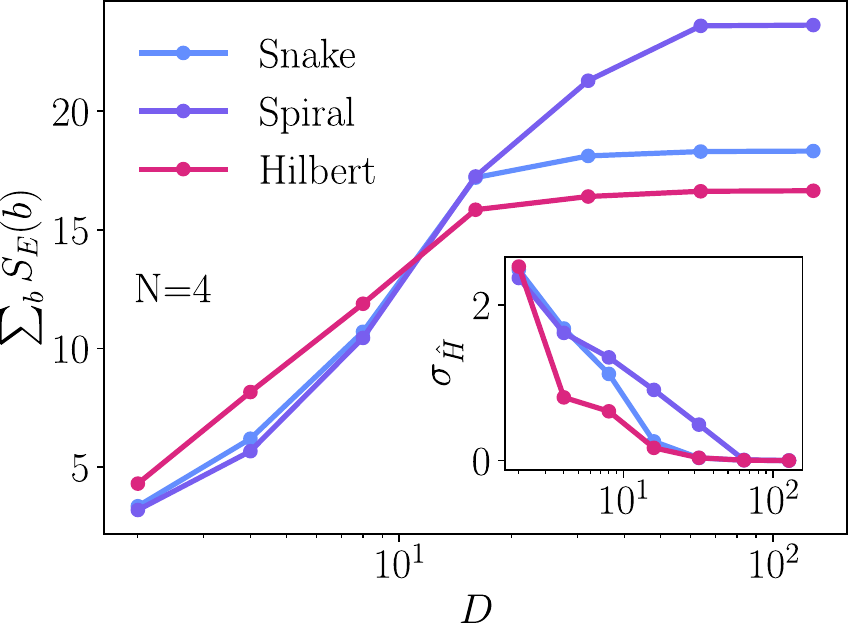}%
}
\caption{%
  \justifying The relative error in the ground state energy against bond dimension \subref{4x4_3paths} and the total bond entanglement entropy \subref{4x4_SE_and_variance} for the snake, spiral, and Hilbert curve paths of the $4\cross4$ SLHAF with OBC. The ED ground state energy is taken as reference. The inset of \subref{4x4_3paths} zooms in on the lower values of $D$. The inset of \subref{4x4_SE_and_variance} shows the variance in ground state energy against bond dimension. }
    \label{4x4_3paths_2fig}
\end{figure}

On the basis of locality metrics alone, using $\delta = 2$ and $q=1$ for $L_{\text{DS}, \delta}$ and $L_{\text{MD},q}$ respectively, we would expect that the snake path will be optimal, with the Hilbert curve and spiral path following in performance respectively. Likewise, if we were to only consider the average number of lattice bonds intersected by a bipartition, the spiral path is clearly a poor choice, as can be seen in Table \ref{3paths_localitytable}. The partitioning line must always pass ``through" the interior of the spiral, which results in a high number of intersected lattice bonds. The snake path and the Hilbert curve have a similar number of intersected bonds. With this is in mind, consider the inset of Fig. \ref{4x4_3paths}, which shows the relative error in the ground state energy per site for the three paths at small bond dimensions, where $\Delta E = \abs{E_0 - E^{\text{ED}}_0}$. Although the spiral path is clearly sub-optimal for these values of $D$, as could be expected \textit{a priori}, the Hilbert curve outperforms the snake path on the $N=4$ SLHAF. This is consistent with the findings of Ref. \cite{Cataldi2021hilbertcurvevs} for the Hilbert curve in the 2D transverse-field quantum Ising model. 

The performance increase of the Hilbert curve over the snake path is not reflected in either the locality metrics nor the total number of bonds cut via bipartitioning. We find that the snake path consistently achieves the lowest average index difference between nearest neighbors of all Hamilton paths on the grid graph up to $N=8$. We suspect this result holds for all $N$, but we are unaware of any such formal proof. It follows that the snake path performs well in locality metrics that equally weights index difference. Notably, the $L_{\text{MD},q}(P)$ metric is able to unequally weight index difference via the parameter $q$. The value of $q$ controls how much large index differences, which translate to long-ranged interactions in the spin chain mapping, are penalized ($q>1$) or tolerated ($q<1$) \cite{MD1986Optimal}. Although there are no known analytic results for the $L_{\text{MD},q}(P)$ metric with $q<1$, Mitchison and Durbin hypothesized that optimal numbering of the grid graph to minimize $L_{\text{MD},q}(P)$ with $q<1$ will be of a fractal nature. Indeed, $L_{\text{MD},q}(P)$ with $q<1$ is the only locality metric that can be lower for the Hilbert curve than for the snake path (see Appendix \ref{Sec:MD}). We may posit that the DMRG algorithm in 2D, at least when applied to the SLHAF, favors paths that minimize $L_{\text{MD},q}(P)$ with $q<1$. Put simply, a handful of very long-ranged interactions are less detrimental to the computational accuracy than many intermediate range interactions. This is precisely the nature of the FASS curves, where within the self-similar motifs the interactions are short-ranged, but between motifs the interactions are very long-ranged. It follows that fractal Hamilton paths will always outperform the snake path for low bond dimensions, as the snake path does not exhibit the clustering properties inherent to the FASS curves. 

\begin{figure}[t]
\centering
\subfloat[\label{Opt2}]{%
  \includegraphics[width=0.24\columnwidth]{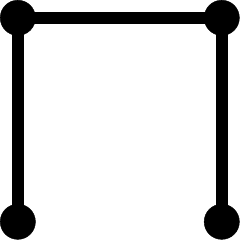}%
}\hfill
\subfloat[\label{Opt3}]{%
  \includegraphics[width=0.24\columnwidth]{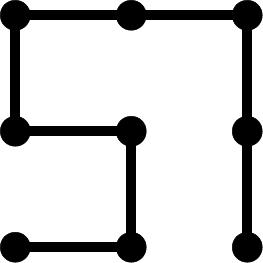}%
}\hfill
\subfloat[\label{Opt4}]{%
  \includegraphics[width=0.24\columnwidth]{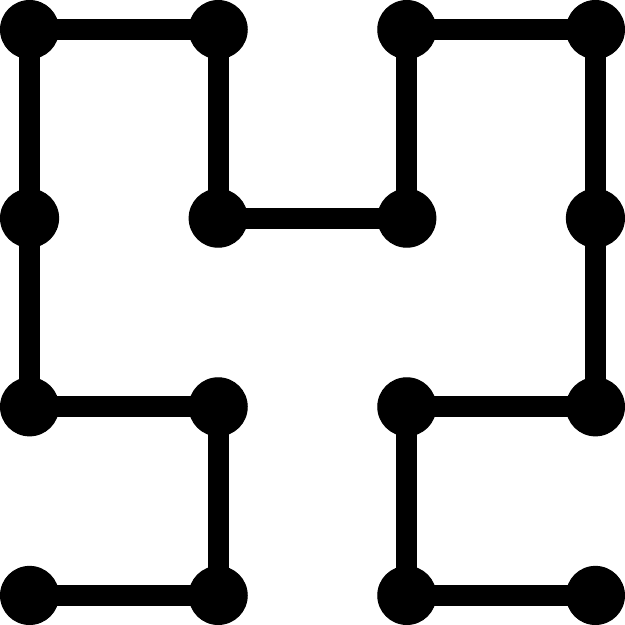}%
}\hfill
\subfloat[\label{Opt5}]{%
  \includegraphics[width=0.24\columnwidth]{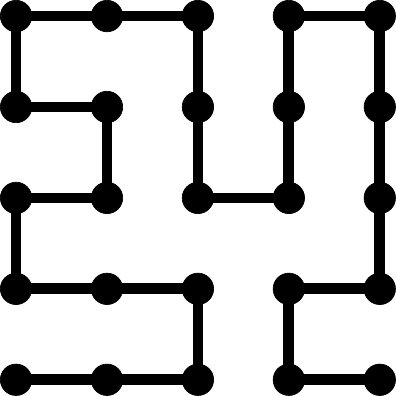}%
}\hfill
\subfloat[\label{Opt6}]{%
  \includegraphics[width=0.31\columnwidth]{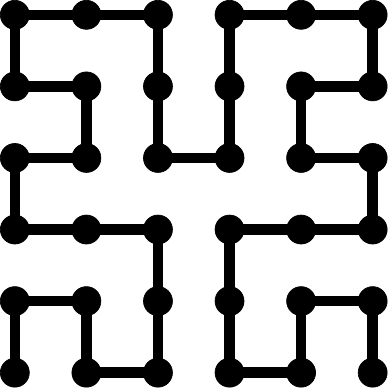}%
}\hfill
\subfloat[\label{Opt7}]{%
  \includegraphics[width=0.31\columnwidth]{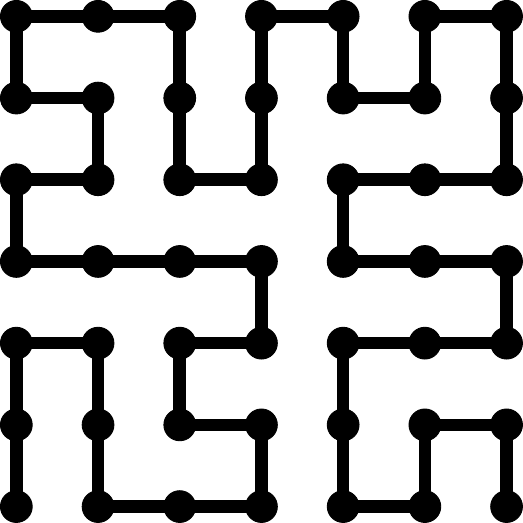}%
}\hfill
\subfloat[\label{Opt8}]{%
  \includegraphics[width=0.31\columnwidth]{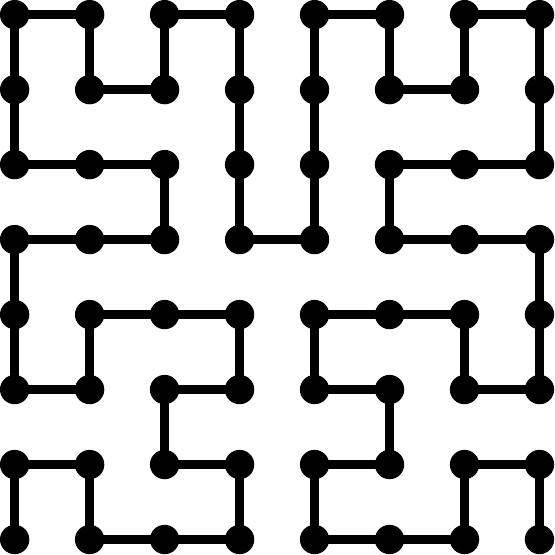}%
}
\caption{%
  \justifying The optimal DMRG paths for the $N\cross N$ SLHAF for $N=2$ \subref{Opt2} through $N=8$ \subref{Opt8}.  }
    \label{optimal_curves}
\end{figure}

For larger bond dimensions, the performance increase of the Hilbert curve becomes less drastic and actually becomes suboptimal to the snake path for $D \gtrsim 30$, as can be seen in Fig. \ref{4x4_3paths}, suggesting that the superior locality preservation of the snake path ultimately becomes appreciable for sufficiently large $D$; however, the performance improvement is well below one order of magnitude at best and is within the expected natural variation of the DMRG algorithm. A more nuanced picture emerges when we consider the total bond entanglement $\sum_b S_E(b)$ and the variance $\sigma_{\hat{H}}$, shown in Fig. \ref{4x4_SE_and_variance} and its inset, respectively. The variance can be thought of as a measure of how accurate the MPS, as given by the DMRG algorithm, approximates the true many-body ground state. Once again, the Hilbert curve is seen to outperform the snake path for $D \lesssim 30$, and above this value they are effectively equivalent. Similar to the variance, the total bond entanglement entropy also quantifies the validity of the algorithm in representing the many-body wavefunction as an MPS, as the required bond dimension generically scales exponentially with $S_E$ in 2D \cite{schollwock2005}. We can see from Fig. \ref{4x4_SE_and_variance} that $\sum_b S_E(b)$ of the Hilbert curve converges to a lower final value than the other paths. Furthermore, it reaches this final value faster. Both of these features provide a much clearer indication of the performance increase of the Hilbert curve; the Hilbert curve MPS more efficiently encodes the entanglement required to approximate the true ground state for a given $D$. 

Although the Hilbert curve converges faster than the snake and spiral paths, we must ascertain the performance of all other possible paths before we are able to conclusively assign optimality to a path. Furthermore, the Hilbert curve is only defined for lattices where $N$ is a power of 2. Other fractal space-filling curves, such as the Peano curve and the Meander curve \cite{Sagan1994SF}, are limited to lattices where $N$ is a power of 3. As such, identifying the motif Hamilton paths that generate FASS curves for prime values of $N > 3$ is highly non-trivial \cite{FASS_paper}. However, such identification is likely worthwhile, as we anticipate that the constructed FASS curves will typically be optimal or very nearly optimal, allowing for an \textit{a priori} informed choice of DMRG path for large systems. 

\begin{figure*}[t]
\centering
\subfloat[\label{3x3_all_paths}]{%
  \includegraphics[width=\columnwidth,trim={0 1.4cm 0 0}]{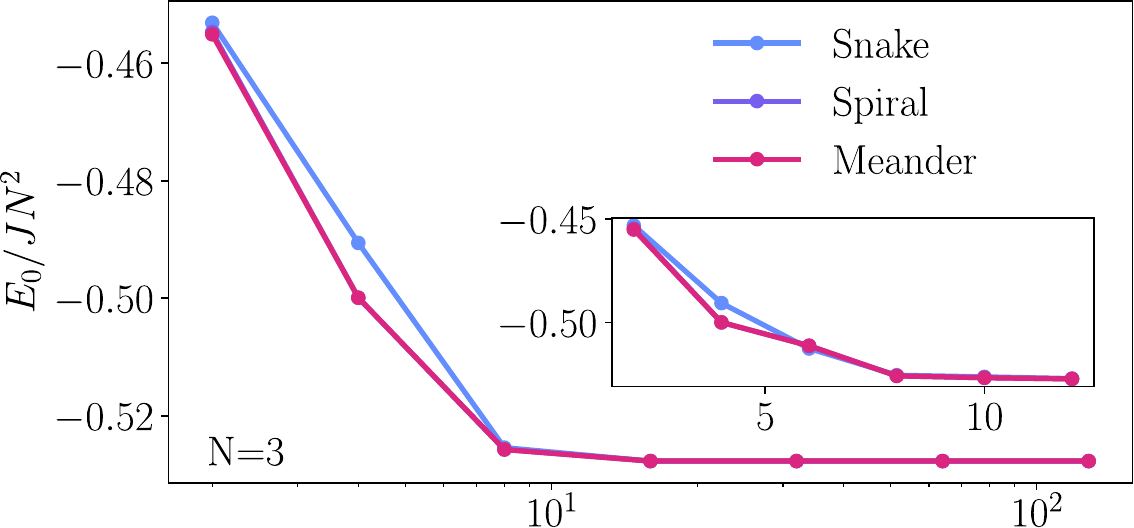}%
}\hfill
\subfloat[\label{4x4_all_paths_log}]{%
  \includegraphics[width=\columnwidth,trim={0 1.4cm 0 0}]{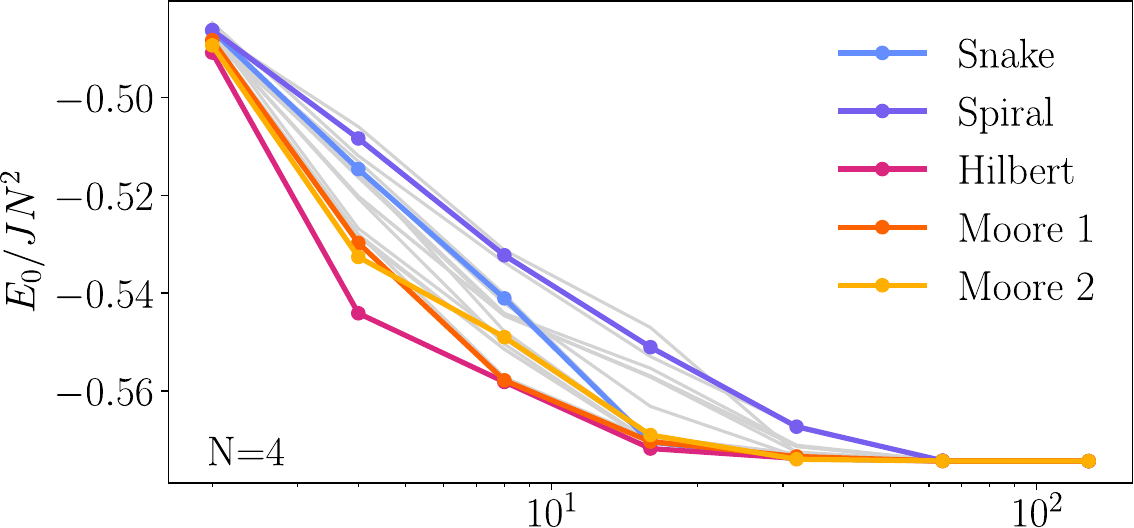}%
}\hfill 
\subfloat[\label{3x3_SE_and_variance}]{%
  \includegraphics[width=\columnwidth]{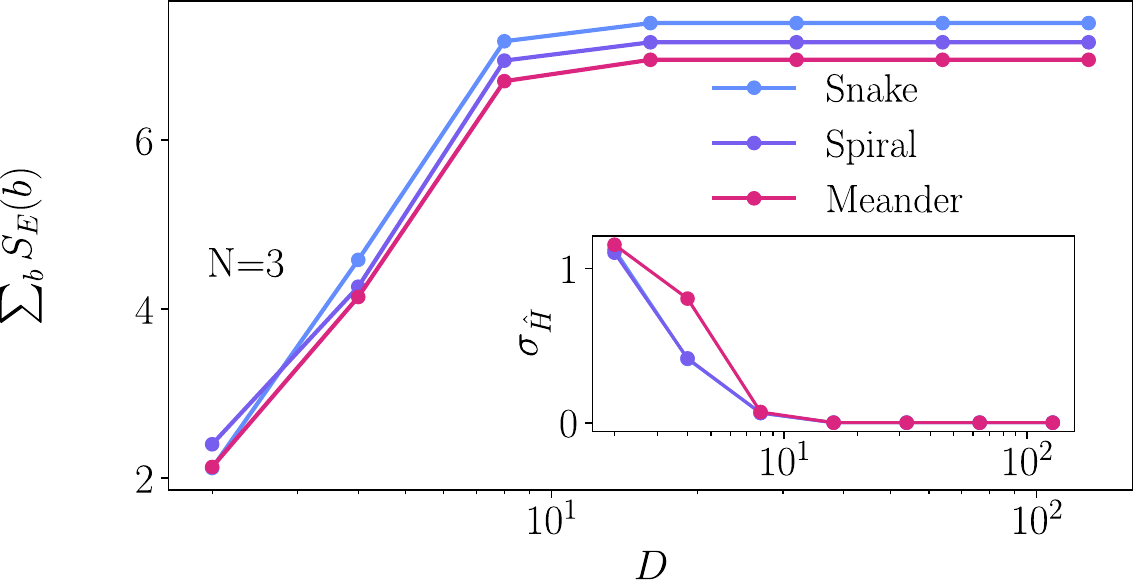}%
}\hfill
\subfloat[\label{4x4_all_SE_and_variance}]{%
  \includegraphics[width=\columnwidth]{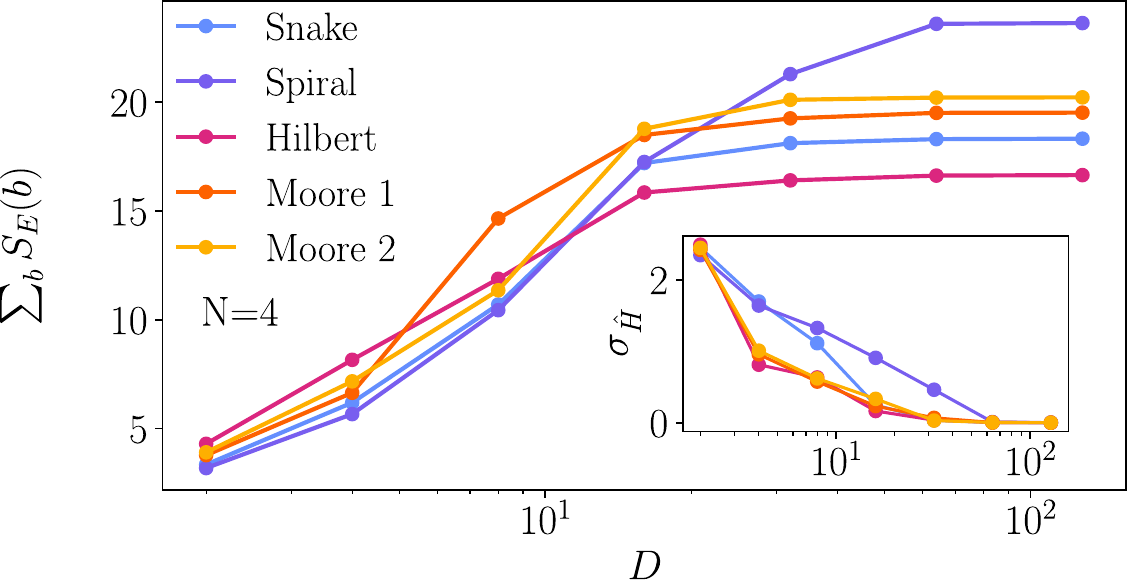}%
}
\caption{%
  \justifying \subref{3x3_all_paths}, \subref{4x4_all_paths_log} The ground state energy per site of the $N=3$ \subref{3x3_all_paths} and $N=4$ \subref{4x4_all_paths_log} SLHAF for all symmetry-reduced, corner starting Hamilton paths against bond dimension $D$. The inset of \subref{3x3_all_paths} zooms in on the lower values of $D$. In \subref{4x4_all_paths_log}, the data of the snake, spiral, Hilbert curve, and Moore curve paths have been coloured. All other paths, deemed generic, have been coloured light grey. \subref{3x3_SE_and_variance}, \subref{4x4_all_SE_and_variance} The total bond entanglement entropy of the named symmetry-reduced, corner starting Hamilton paths for the $N=3$ \subref{3x3_SE_and_variance} and $N=4$ \subref{4x4_all_SE_and_variance} SLHAF. Inset is the variance in the ground state energy for these same paths.}
\label{3x3_4x4_2fig}
\end{figure*}

\section{Results}\label{Results}

The following section details our investigation into path optimality for the 2D DMRG algorithm. We discuss our computational methodology in Appendix \ref{Comp_dets}. In Fig. \ref{optimal_curves} we show our claimed optimal paths for the $N \cross N$ grid graph up to $N = 8$, which we justify in the following subsections. The optimal paths for $N= 2,3,5,$ and $7$ are equivalently FASS motifs that can be used to construct high-performing paths for larger values of $N$. 

\subsection{$\mathbf{3 \cross 3}$}

There are three symmetry-reduced Hamilton paths of the $N=3$ grid graph. We denote the three paths as the snake path, the spiral path, and the `meander' path, in reference to the motif of the Peano-meander curve (see Fig. \ref{Meander_motif}) \cite{Sagan1994SF}. It is notable that the $N=3$ snake path is the motif of the standard Peano curve and thus they are identical for this system size. For $N=9$ the snake path and the Peano curve become distinct. The ground state energy per site of the $3 \cross 3$ SLHAF with OBC is shown in Fig. \ref{3x3_all_paths}. For bond dimension values of $D>12$ the ground state energy per site quickly converges to the value of $E^{\text{ED}}_0/JN^2 \approx -0.5277$ as given by ED, and little difference can be seen between the three paths. 

The spiral and the meander paths yield the same ground state energy to within three decimal places and are largely indistinguishable for all bond dimensions. $N=3$ is the only lattice size for which the spiral path is close to optimal, which suggests that $N=3$ is too small for the long distance interactions inherent to the spiral path to become detrimental. For $D=6$ we find the snake path transiently becomes optimal, but returns to suboptimal for all other values of $D$. This points to the inherent variability of results from DMRG, and their sensitivity to simulation parameters. The variance also shows very little difference between the three paths except at low $D$ where the meander path performs slightly worse. However, the meander path does show an overall lower value of the total bond entanglement entropy in the converged limit. 

As such, although it is difficult to conclusively say that the meander path is the optimal path  for $N=3$ based solely on the goundstate energy convergence, the spiral path is unable to be FASS tiled and has a slightly higher converged entanglement entropy. For these reasons we denote the meander curve as the optimal Hamilton path for $N=3$. 
\subsection{$\mathbf{4 \cross 4}$}

\begin{figure*}[t]
\centering
\subfloat[\label{5x5_all_logplot}]{%
  \includegraphics[width=\columnwidth,trim={0 1.4cm 0 0}]{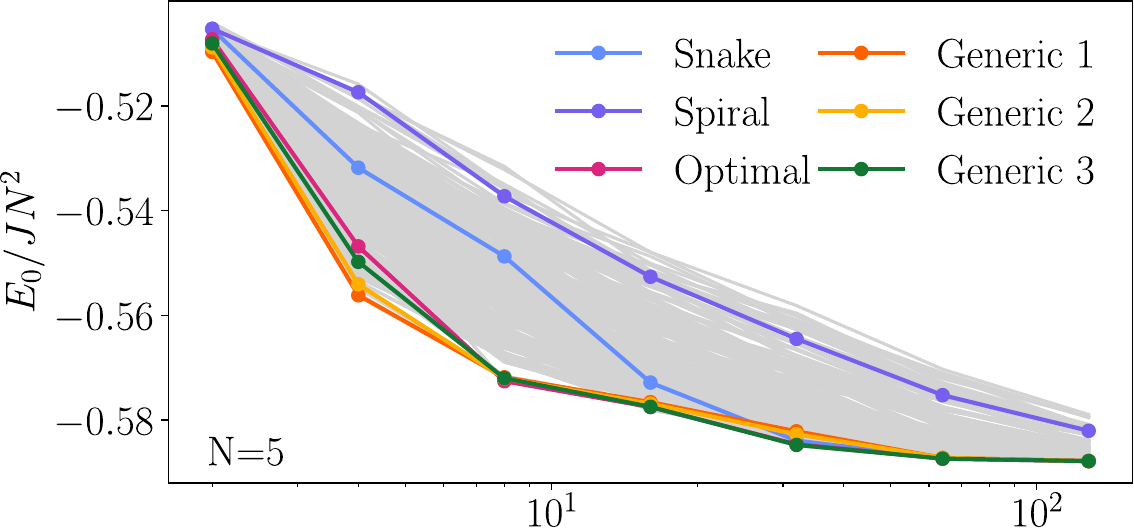}%
}\hfill
\subfloat[\label{6x6_all_logplot}]{%
  \includegraphics[width=\columnwidth,trim={0 1.4cm 0 0}]{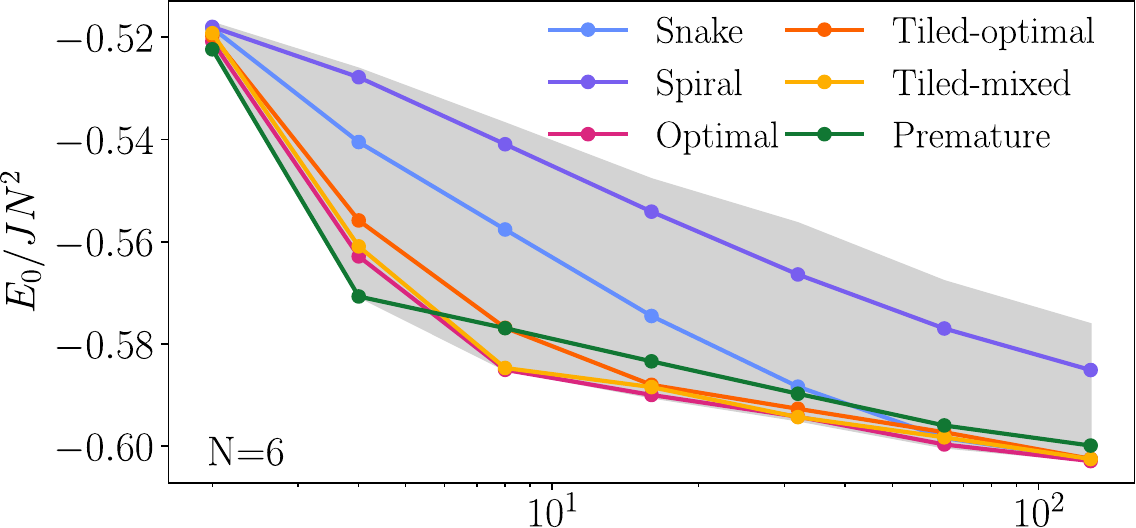}%
}\hfill
\subfloat[\label{5x5_SE_and_variance}]{%
  \includegraphics[width=\columnwidth]{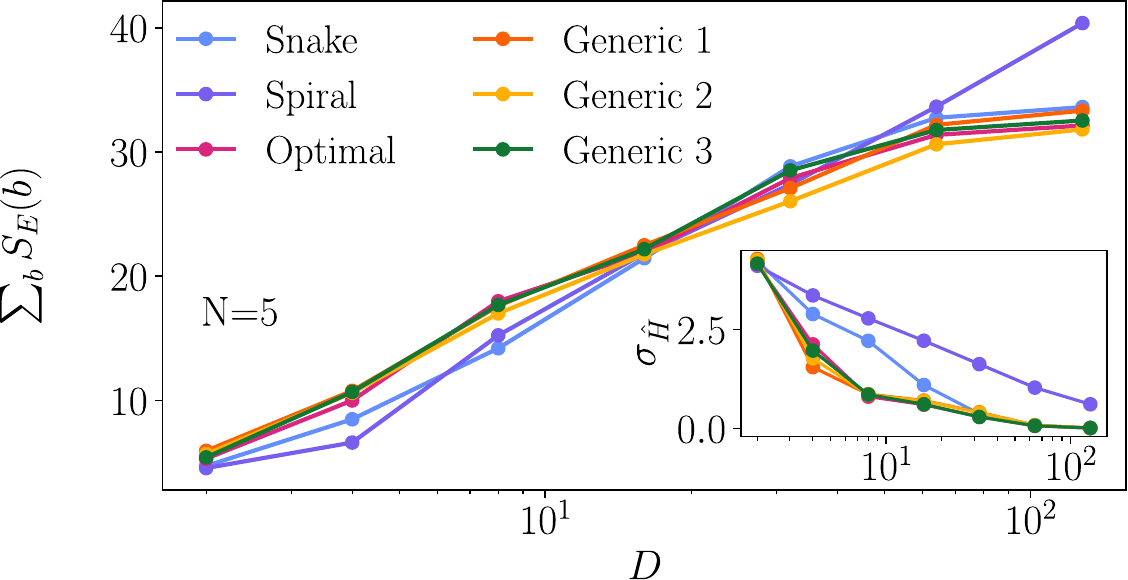}%
}\hfill
\subfloat[\label{6x6_SE_and_variance}]{%
  \includegraphics[width=\columnwidth]{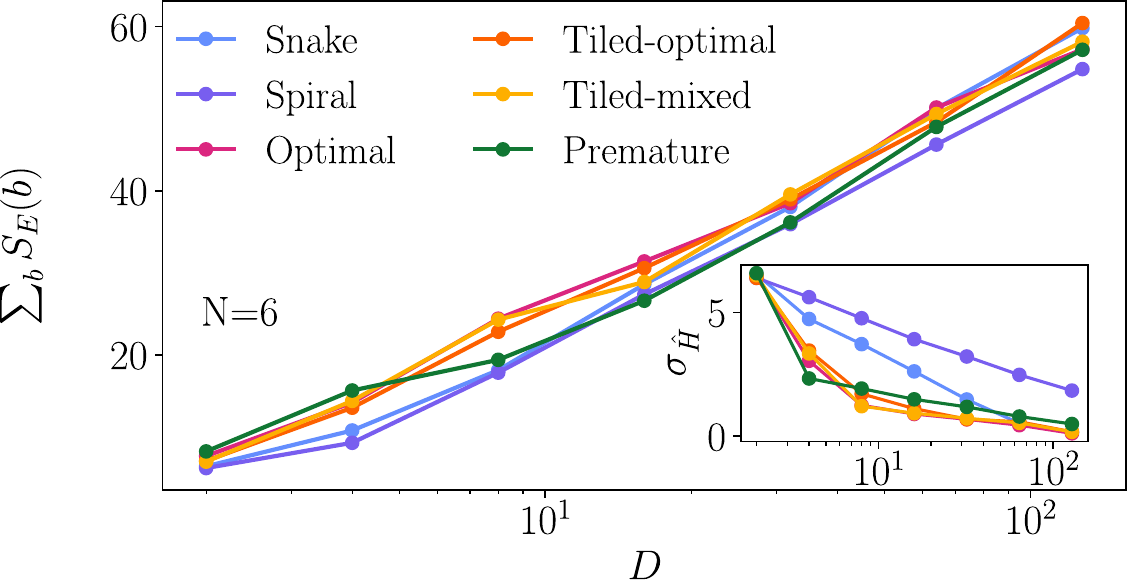}%
}
\caption{%
\justifying \subref{5x5_all_logplot}, \subref{6x6_all_logplot} The ground state energy per site of the $N=5$ \subref{5x5_all_logplot} and $N=6$ \subref{6x6_all_logplot} SLHAF for all symmetry-reduced, corner starting Hamilton paths against bond dimension $D$. In \subref{5x5_all_logplot}, the data of the snake, spiral, Optimal, and three generic paths have been coloured. All other paths have been coloured light grey. In \subref{6x6_all_logplot}, the data of the snake, spiral, Optimal, the worst performing tiled-optimal, the best performing tiled-mixed, and the premature paths have been coloured. The data of all other paths are enclosed in the grey polygon. \subref{5x5_SE_and_variance}, \subref{6x6_SE_and_variance} The total bond entanglement entropy of the named symmetry-reduced, corner starting Hamilton paths for the $N=5$ \subref{5x5_SE_and_variance} and $N=6$ \subref{6x6_SE_and_variance} SLHAF. Inset is the variance in the ground state energy for these same paths.
}
  \label{5x5_6x6_2fig}
\end{figure*}
As mentioned in Sec. \ref{PIDMRG} there are 23 symmetry-reduced, corner starting Hamilton paths through the $N=4$ grid graph. Figure \ref{4x4_all_paths_log} shows the ground state energy per site for all such paths against bond dimension. Generic paths that are suboptimal, asymmetric, or not geometrically well defined (e.g. the spiral path is suboptimal but well defined), are coloured light grey. The majority of these paths are slight variants of the named paths; variants of the spiral path typically perform poorly, and variants of the Hilbert curve perform relatively well. Besides the three paths introduced in Sec. \ref{PIDMRG} there exists a Hamilton cycle variant of the Hilbert curve known as the Moore Curve \cite{Sagan1994SF}. As we are only considering corner starting Hamilton paths, there will always be two paths corresponding to the removal of either the vertical or horizontal bond at the corner of the Hamilton cycles. These two paths, denoted `Moore 1' and `Moore 2', are shown in Appendix \ref{App_paths}. The Moore curve performs very well compared to all other curves, as could be expected from its relation to the Hilbert curve. The optimal path is indeed confirmed to be the Hilbert curve, as most clearly seen from the total bond entanglement entropy and variance data shown in Fig. \ref{4x4_all_SE_and_variance}. 

\subsection{$\mathbf{5 \cross 5}$}

For $N=5$ we start to see the difficulty caused by the rapid growth of path number with respect to lattice size. The snake and spiral paths are the only two paths that are predefined, with the other 345 symmetry-reduced, corner starting paths unable to be easily described. Of these, 43 paths are FASS para-tile motifs. Figure \ref{5x5_all_logplot} shows the ground state energy against bond dimension for all 347 paths, where once again we have coloured the generic paths grey, now with three exceptions which are shown in Appendix \ref{App_paths} and denoted Generic paths 1,2, and 3 respectively. Generic paths 1 and 2 outperform all other paths besides the optimal path for $D = 4$ and $D = 8$ respectively. Generic path 3 is a variant of the optimal path that performs surprisingly well, even outperforming the optimal at times. We include these generic paths to highlight the difficulty with identifying a single optimal path among many paths, especially for odd numbered $N$. No one path significantly outperforms all other paths. Indeed, unlike the $N=4$ case, the path that yielded the lowest ground state energy is different for 6 of the 7 bond dimensions tested. 

With this difficulty in mind, we denote the path shown in Fig. \ref{Opt5} as the optimal path for the $N=5$ grid graph. This path is unambiguously optimal among the 43 possible FASS motifs and was briefly overall optimal for $D = 8$. It is noteworthy that this path is reminiscent of the Hilbert curve, as can be seen by considering its bottom right $4 \cross 4$ square. Fig. \ref{5x5_SE_and_variance} and its inset show the total entanglement entropy and the variance of the six named paths for $N=5$, respectively. Except for very small values of $D$ the optimal path and its relative Generic 3 achieve the lowest variance values of the named paths. With respect to the total bond entanglement entropy, it is clear that all paths have not reached convergence at $D=128$. Notably, all the named paths besides the spiral path are showing the beginnings of a plateau in entropy, with the optimal path showing the smallest increase in total $S_E$ between $D=64$ and $D=128$. This further supports our assertion that we have identified the optimal path for the $N=5$ SLHAF. 

\subsection{$\mathbf{6 \cross 6}$} \label{6x6_section}

For $N=6$ the number of generic paths becomes untenable, and we do not plot all individual ground states energies for every path in Fig. \ref{6x6_all_logplot}. Instead, the light grey polygon encloses the space delineated by the maximal and minimal $E_0/JN^2$ values of the generic paths for each bond dimension; therefore, the values of all paths except the named paths are guaranteed to fall within this polygon. $N=6$ is also the first lattice size that we may introduce what we term as `tiled-optimal', `tiled-mixed', and `premature' paths. A tiled-optimal path is an FASS tiling of an optimal path of a smaller $N$ lattice. The first such tiled-optimal path is by definition the second order Hilbert curve, as there is only one $N=2$ grid graph path, which is therefore optimal. One of the tiled-optimal paths constructed by Hilbert tiling the Meander curve, as discussed in Sec. \ref{Ham_paths_Curves}, is shown in Appendix \ref{App_paths}. For brevity, in this work we shall always imply the Hilbert tiling when referring to the tiled-optimal paths unless otherwise stated. A tiled-mixed path is a tiling of any smaller lattice paths that are not necessarily optimal; the path corresponding to a tiling of two snake paths and two meander paths can be seen in Appendix \ref{App_paths}. Note that since the $N=3$ snake and meander paths are ortho- and para-tiles respectively, this tiling is not technically a Hilbert tiling. Lastly, a premature path is one that performs exceptionally well for very low bond dimensions, typically $D=2$ or $D=4$, but is ultimately suboptimal for larger bond dimensions. The best performing premature path for $N=6$ is shown in Appendix \ref{App_paths}. 

The $N=6$ tiled-optimal paths are constructed from the meander curve, which is asymmetric. There are ten unique ways that an asymmetric motif can be Hilbert tiled, only four of which result in paths that are symmetric. We find that one of the symmetric tiled-optimal paths is the overall optimal path for the $N=6$ grid graph, as shown in Fig. \ref{Opt6}. The ground state energy per site of the worst performing tiled-optimal path is plotted in Fig. \ref{6x6_all_logplot} in orange. Although it is an improvement over the snake path, it is comfortably outperformed by the optimal path, which reflects the sensitivity of the DMRG algorithm to the chosen path; these two tiled-optimal paths differ only by a reflection of their two upper quadrants (see Fig. \ref{Opt6} and Appendix \ref{App_paths}). We find that paths whose quadrants are identifiably $N=3$ paths, such as the $N=6$ tiled-mixed path shown in Appendix \ref{App_paths}, almost always outperform generic paths. This strongly suggests that the performance of 2D DMRG paths is broadly improved for paths that demonstrate self-similarity, and further improves if the self-similar partitions are themselves optimal. In Fig. \ref{6x6_SE_and_variance} we see the same behavior as for the previous lattice sizes. Again, we find that variance performance generally correlates with ground state energy accuracy, and that none of the paths have reached convergence for $D=128$. We do identify that the early performance of the premature path is indicated by its large total entanglement entropy for small $D$, which is then outpaced by the tiled-optimal paths. 
\subsection{$7 \cross 7$}

\begin{figure}
\centering
\subfloat[\label{7x7_all_logplot}]{%
  \includegraphics[width=\columnwidth,trim={0 1.4cm 0 0}]{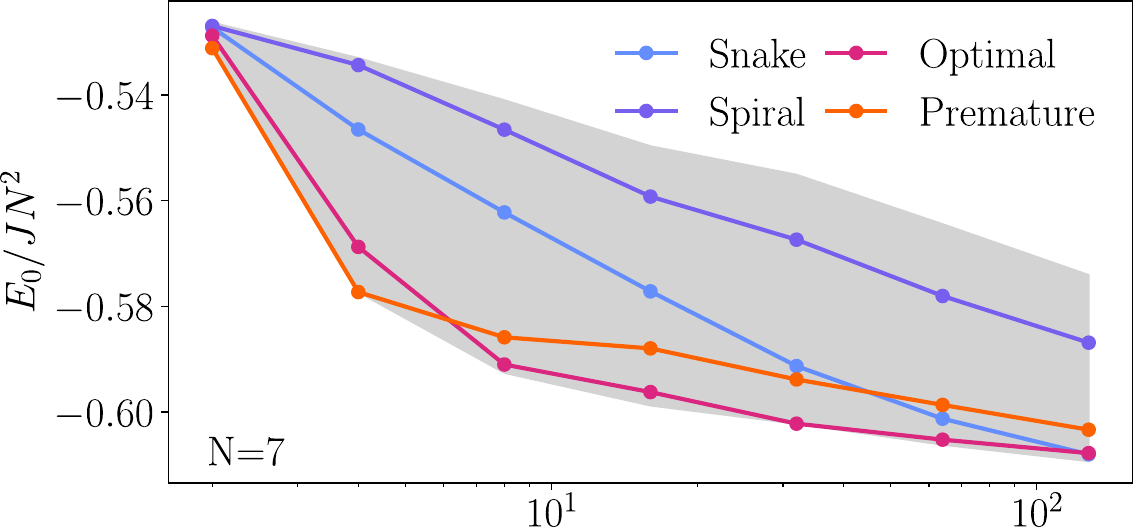}%
}\hfill
\subfloat[\label{8x8_all_logplot}]{%
  \includegraphics[width=\columnwidth,trim={0 0 0 0}]{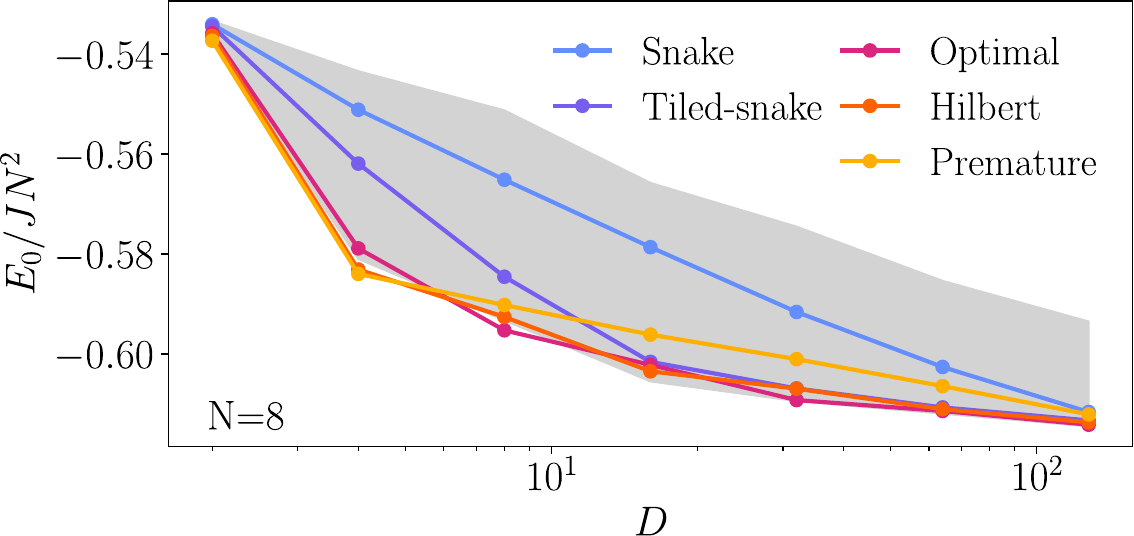}%
}
\caption{%
  \justifying The ground state energy per site of the N = 7 SLHAF for all symmetry-reduced, corner starting Hamilton paths against bond dimension D. Data of the snake, spiral, optimal and premature paths have been coloured. The data of all other paths fall within the bounds of the grey polygon. The ground state energy per site of the N = 8 SLHAF for the 888 symmetric FASS motif paths against bond dimension $D$. Data for the snake, tiled-snake, optimal, Hilbert and premature paths have been coloured. The data of the remaining paths fall within the bounds of the grey polygon.}
  \label{7x7_8x8_2fig}
\end{figure}

In Fig. \ref{7x7_all_logplot} we once again show the polygon enclosing the maximal and minimal energy values of the generic paths, and we colour the values of the snake, spiral, optimal, and premature paths, for the $N=7$ SLHAF. The optimal path is once again identified as the best performing path among the FASS motifs and is shown in Fig. \ref{Opt7}. Surprisingly, for $D=32$ it outperforms all other paths, but is typically only within the top 100 best performing paths for all other values of $D$, owing to the numeric instability inherent to larger values of $N$. The premature path is shown in Appendix \ref{App_paths}. We also show the total bond entanglement entropy and variance for $N=7$ and $N=8$ in Appendix \ref{7x7_8x8_appendix} as the trends are not manifestly different from those seen for $N=6$ and $N=5$. 

\subsection{$8 \cross 8$}

For $N=8$ there are close to 86 million symmetry-reduced, corner starting Hamilton paths, of which 4.4 million are valid FASS motifs. We can make a informed search by noting that the optimal paths for $N=4$ and $N=6$ are symmetric, and we posit that the optimal FASS motifs for even $N$ are likely to follow this pattern. There are 888 symmetric FASS motifs for the $N=8$ grid graph; the Hilbert curve and the snake path are among this group. As such, we cannot definitively say that the overall optimal path will be within this subset, but it is likely given the previous lattice sizes. In Fig. \ref{8x8_all_logplot} we identify 5 paths among these symmetric FASS motifs: the snake, the Hilbert tiled snake, the optimal (see Fig. \ref{Opt8}), the Hilbert curve, and the premature path. The Hilbert tiled snake and premature paths can be seen in Appendix \ref{App_paths}.

The $N=8$ optimal path is a tiled-mixed path with a very similar structure to the Hilbert curve, with the two upper quadrants replacing the Hilbert motif with an assymetric $4 \cross 4$ motif. Although both exhibit similar levels of performance, the optimal path is overall best for $D=8$ and $D=32$, whereas the Hilbert curve is never overall optimal. The Hilbert tiled snake path shows very good performance in comparison to the standard snake mapping. As the snake path is always constructible for any value of $N$, this suggests that an FASS tiling of the snake path is always a better choice for ground state searches than the snake path itself; however, it is still outperformed by path informed DMRG using the tiled-optimal paths. As such, the first value of $N$ that the tiled snake path would be useful is $N=22$, as determination of the optimal FASS motif for $N=11$ is likely impossible (see below). 

Ultimately, although the Hilbert curve was slightly outperformed by the overall optimal path, we have shown that a Hilbert tiling of an optimal path on the $N \cross N$ lattice yields the optimal or very close to the optimal path for the $2N \cross 2N$ lattice. As such, for ease of construction it is not unreasonable to prescribe the Hilbert curve as the optimal path for grid graphs with $N$ being a power of 2. Furthermore, this suggests that Hilbert tiling of prime numbered FASS motifs will always yield high performance paths, and we therefore can prescribe high-performing paths for all $N = p\cdot2^n \quad \forall \quad n \, \in \, \mathbb{Z}+$ with prime $p \leq 7$. 

\section{Constructing high-performing paths for large systems}

As was most clearly seen in Sec. \ref{6x6_section} for the $N=6$ SLHAF, we are able to construct high-performing DMRG paths by FASS tiling smaller optimal paths. We claim that such paths for $N\geq8$ can only ever be deemed `high-performing', as exhaustive determination of optimality is not currently feasible with modern computational resources. Even for $N=8$ in this investigation we only assert the optimal path among the symmetric FASS motifs; the so called `palindromic' para-tiles in Ref. \cite{FASS_paper}. For $N=9$ there are around 1 billion FASS motifs \cite{oeisA384173}. As such, for larger values of $N$ we are only able to assess the performance of constructible paths, such as the snake, the spiral, and tiled paths. 

\begin{figure}[ht]
    \includegraphics[width=\columnwidth]{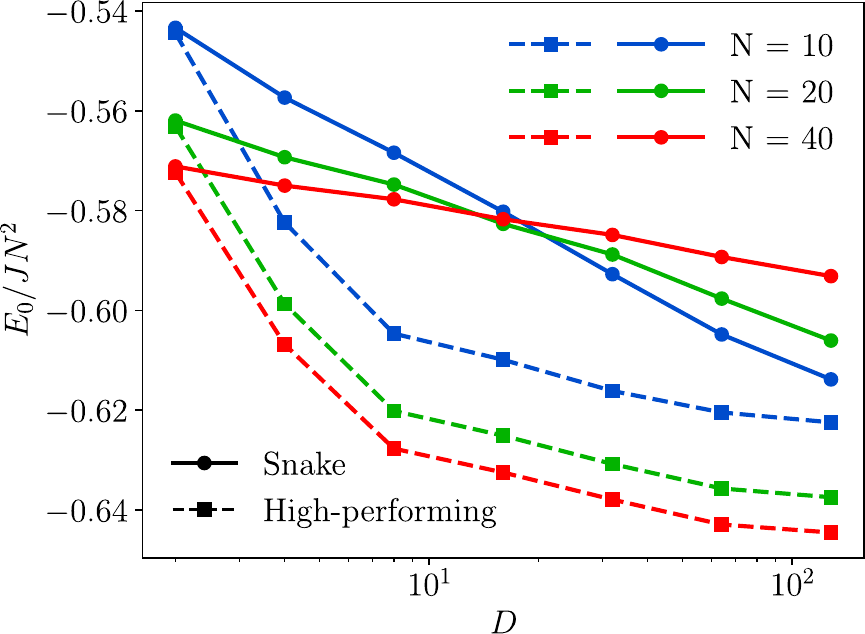}
	\caption{The ground state energy per site for the \hbox{$N=10, \, 20,$ and $40$} SLHAF against bond dimension $D$ for the snake and optimal-tiled high-performing paths.}
	\label{10_20_40_fig}
\end{figure}

As a proof of concept of the capabilities of path-informed DMRG let us consider the SLHAF for $N=10,20,$ and $40$ with OBC. These system sizes are unreachable by ED and can be difficult to simulate using 2D DMRG with intermediately sized values of bond dimension $D$. Furthermore, no well-known space-filling curves, such as the Hilbert curve or the Peano curve, exist for these sizes. As such, calculation on such sizes using 2D DMRG typically utilizes snake path mapping. Here, we instead use the Hilbert tilings of the $N=5$ optimal path and compare it to the results given by the snake path. When Hilbert tiling the $N=5$ optimal motif there are four symmetric resultant curves. The best performing tiled optimal path for $N=10$ is shown in Appendix \ref{App_paths}. This curve is then Hilbert-tiled to generate the $N=20$ and $N=40$ high-performing paths. Figure \ref{10_20_40_fig} shows the ground state energy per site of the $N=10, 20,$ and $40$ SLHAF for the snake and high-performing paths. 

We see that for all three values of $N$, the ground state energy of the high-performing paths is lower than the snake path at all tested bond dimensions. For the $N=40$ lattice the relative difference in ground state energy between the two paths is up to approximately $9\%$. This accuracy increase is most readily apparent for relatively small bond dimensions, but even more striking is the perceived scaling of the performance enhancement: the high-performing paths show consistent, near-exponential scaling performance against bond dimension. This indicates a significant resource boost in accuracy per unit memory, particularly for large system sizes. The consistent performance scaling is an example of the `scale invariance' inherent to fractal shapes, whereby physical properties remain unchanged under dilations due to their self-similarity. This is a key advantage of performing path-informed DMRG over the snake mapping for larger values of $N$, as the snake mapping is not scale invariant and becomes systematically worse as system size increases.

\section{Conclusions}

In this work we have shown the performance increase that is possible from an informed choice of mapping when using 2D DMRG to study the square lattice Heisenberg model. We find that path mappings corresponding to the fractal space-filling curves achieve much faster convergence in ground state searches than the commonly utilized snake path mapping. The FASS paths exhibit a more efficient scaling and overall lower value of both total bond entanglement entropy and variance in ground state energy. We investigated locality metrics to quantify the potential origins of the performance increase of the FASS paths. We postulate that the 2D DMRG algorithm favours optimal mappings of the Mitchison and Durbin locality metric with $q<1$, which agrees with their hypothesis that such mappings would have a fractal nature. Ultimately, we present the optimal paths of the FASS motifs for SLHAF up to $N=8$ and use these motifs to construct larger FASS paths that we claim are high-performing. 

The performance increase is largely restricted to the values obtained before convergence is reached at large bond dimensions. As such, path-informed DMRG is not, nor does it attempt to be, an improvement on 2D DMRG using the snake mapping for system sizes where convergence can be reached for intermediately sized values of $D$. The strength of path-informed DMRG is exemplified for \textit{larger systems} where convergence is not easily obtained. In these cases, regardless of how over-truncated the resultant MPS will be, we guarantee that using an FASS path will achieve a more accurate approximation of the many-body ground state of the SLHAF.

The optimal FASS motifs presented in this work allow the construction of high-performing paths for about half of the composite values of $N$ under $50$. For prime values of $N$ above $7$ and all resulting composite values, we cannot prescribe a high-performing path. Furthermore, we suspect that finding the optimal FASS motif for $N=11$ among approximately $5\cross10^{14}$ paths \cite{oeisA384173}, through exhaustive search, is likely beyond modern computational means. However, there are many approximate space-filling curve schemes in the field of data-analysis for arbitrary values of $N$ which are likely sufficient to perform path-informed DMRG \cite{Bohm2021FurHilbert}. 

Throughout this investigation, we have restricted the scope to only considering open boundary conditions for the SLHAF. A natural extension is to consider both cylindrical and full periodic boundary conditions and the effect this has on path optimality. Furthermore, inclusion of additional terms in the Heisenberg Hamiltonian such as anisotropy, next-nearest-neighbor coupling, antisymmetric exchange, and magnetic fields are expected to strongly affect the relative performance of different paths, since the dynamics will be favored towards the anisotropy or added magnetic fields \cite{PhysRevE.108.044128}. The Hamilton paths on the square grid graph are equally applicable to the triangular lattice, and we would naturally anticipate that the FASS paths will again outperform the snake mapping for the triangular Heisenberg model. However, there are other Hamilton paths on the triangular lattice beyond those it shares with the square lattice; consequently, the determination of the optimal paths on the triangular lattice, as well as many other lattices, remains an open question.

\begin{acknowledgments}

We thank A. Howroyd, I. Iakoupov, D. Poletti, and T. Pollinger for helpful conversations. All computations were performed with the Deigo cluster at the Okinawa Institute of Science and Technology, and we acknowledge the Scientific Computing and Data Analysis Section at OIST. The generated codes and data are freely available from Ref. \cite{snakeGitHub}. 
\end{acknowledgments}

\begin{appendix}
\section{Mitchison and Durbin locality metric of the Snake path and the Hilbert curve}\label{Sec:MD}

\begin{figure}[!htb]
\centering
\subfloat[\label{MD_16x16}]{%
\includegraphics[width=0.49\columnwidth]{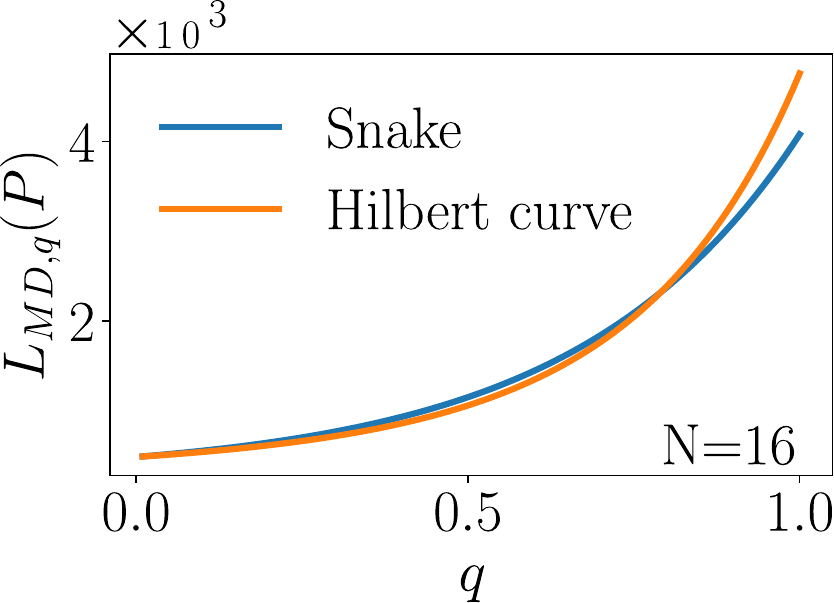}%
}\hfill
\subfloat[\label{MD_64x64}]{%
  \includegraphics[width=0.49\columnwidth]{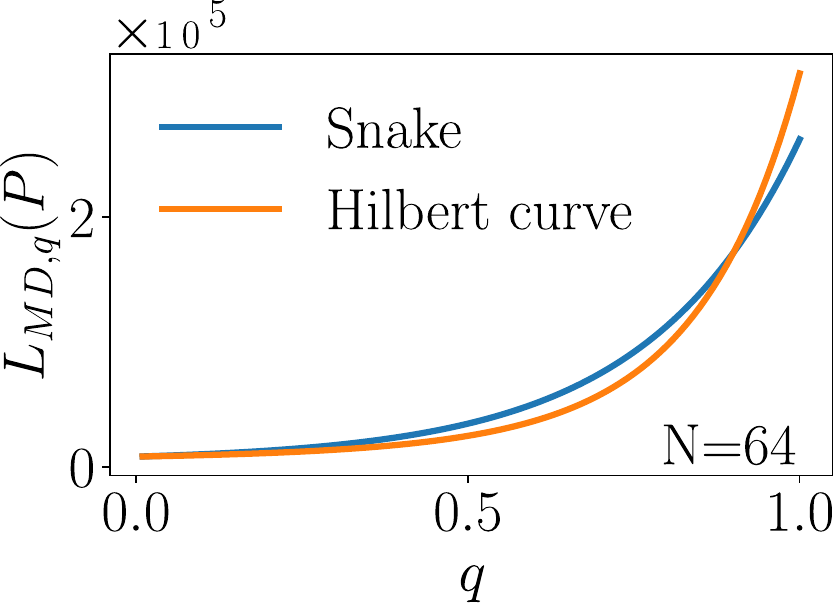}%
}
\caption{%
  \justifying   The Mitchison and Durbin locality metric, $L_{\text{MD},q}(P)$, for the snake path and the Hilbert curve on the \subref{MD_16x16} $N=16$ and \subref{MD_64x64} $N=64$ grid graph.  }
    \label{MD_plots}
\end{figure}

Figure \ref{MD_plots} shows the $L_{\text{MD},q}(P)$ locality metric for the snake path and the Hilbert curve on the $N=16$ and $N=64$ grid graphs. For $q=1$ the snake path is found to always yield a smaller locality metric; for $q<1$ there exists a crossover point at which the Hilbert curve yields a smaller locality metric. The crossover value of $q$ scales with system size and approaches unity for $N \rightarrow \infty$. 

\begin{figure}[t]
\centering
\subfloat[\label{Moore1}]{%
  \includegraphics[width=0.18\columnwidth]{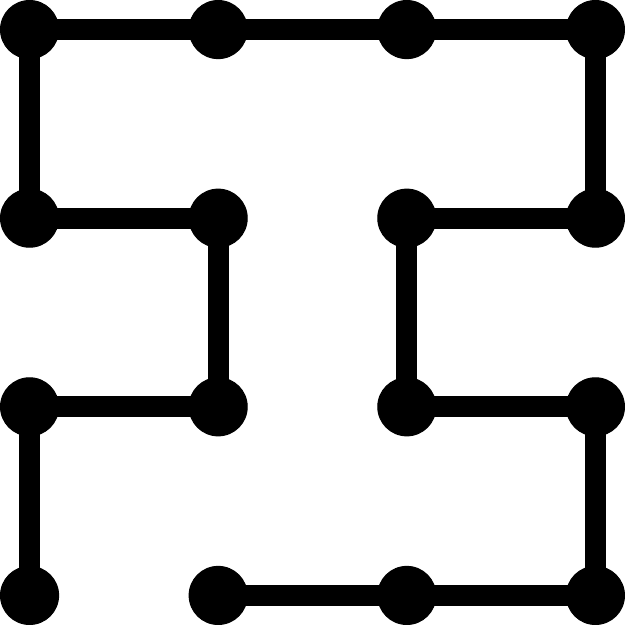}%
}\hfill
\subfloat[\label{Moore2}]{%
  \includegraphics[width=0.18\columnwidth]{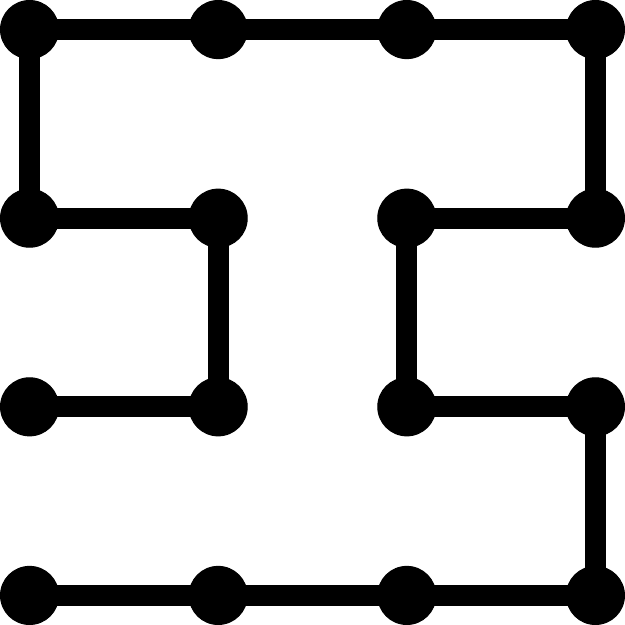}%
}\hfill
\subfloat[\label{5x5Generic1}]{%
  \includegraphics[width=0.18\columnwidth]{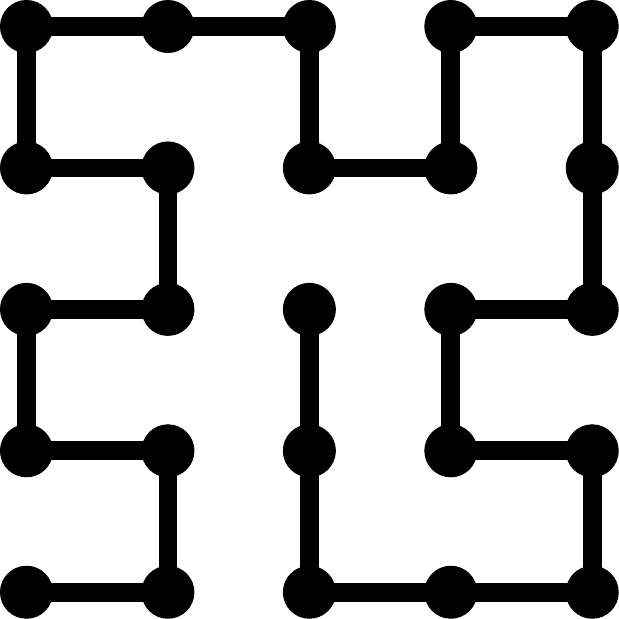}%
}\hfill
\subfloat[\label{5x5Generic2}]{%
  \includegraphics[width=0.18\columnwidth]{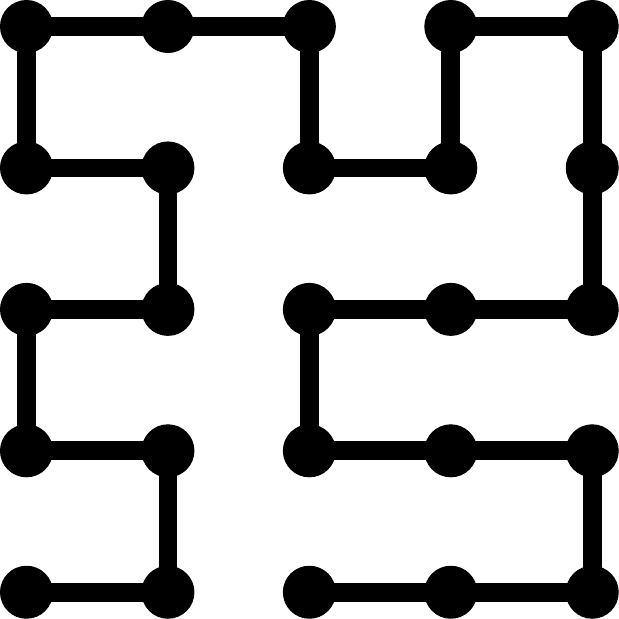}%
}\hfill
\subfloat[\label{5x5Generic3}]{%
  \includegraphics[width=0.18\columnwidth]{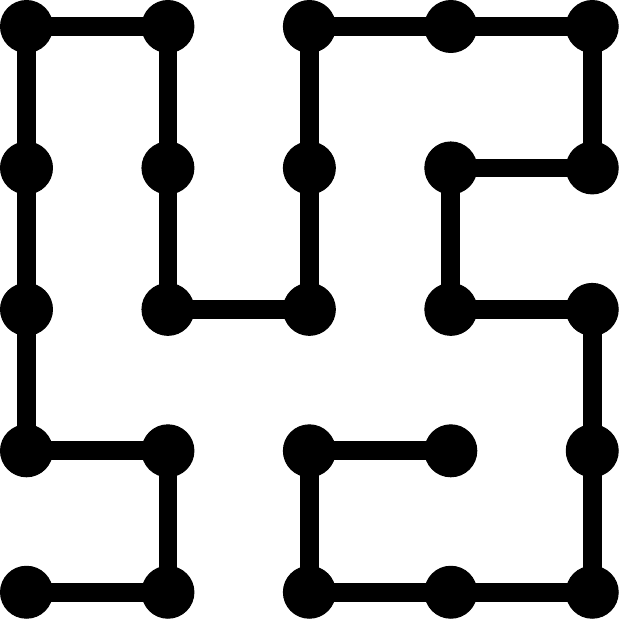}%
}\hfill
\subfloat[\label{6x6WorstTiledOpt}]{%
  \includegraphics[width=0.3\columnwidth]{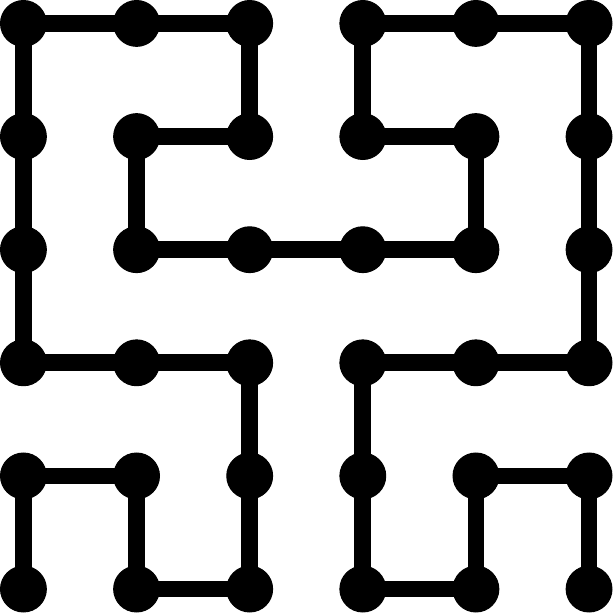}%
}\hfill
\subfloat[\label{6x6_TiledMixed}]{%
  \includegraphics[width=0.3\columnwidth]{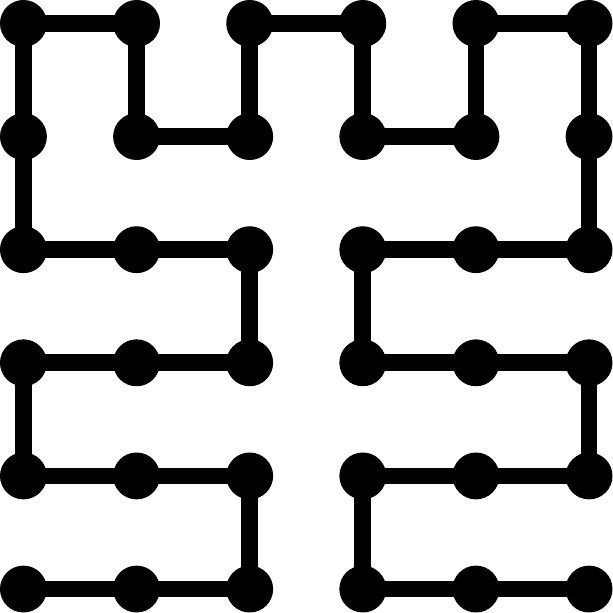}%
}\hfill
\subfloat[\label{6x6Premature}]{%
  \includegraphics[width=0.3\columnwidth]{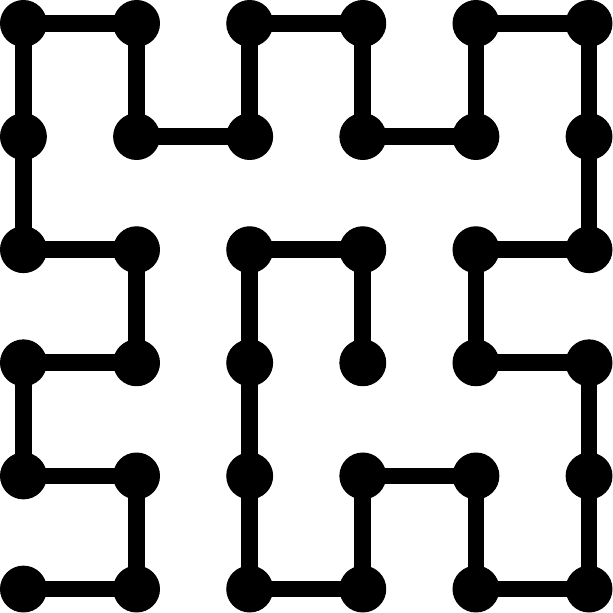}%
}\hfill
\subfloat[\label{7x7Premature}]{%
  \includegraphics[width=0.46\columnwidth]{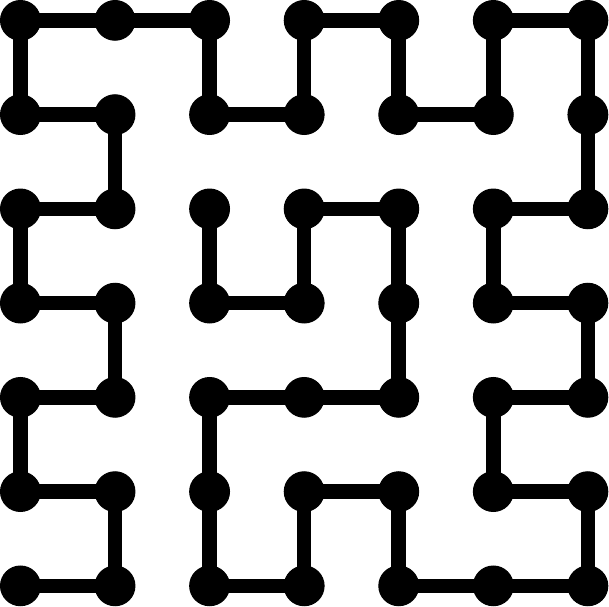}%
}\hfill
\subfloat[\label{8x8TiledSnake}]{%
  \includegraphics[width=0.46\columnwidth]{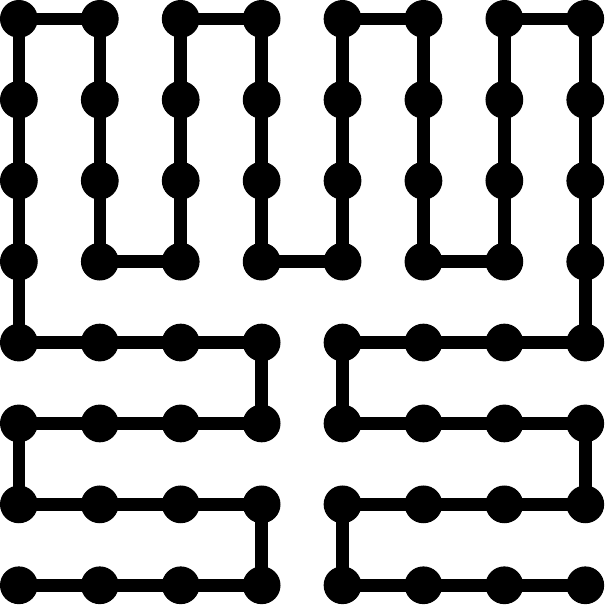}%
}\hfill
\subfloat[\label{8x8Premature}]{%
  \includegraphics[width=0.46\columnwidth]{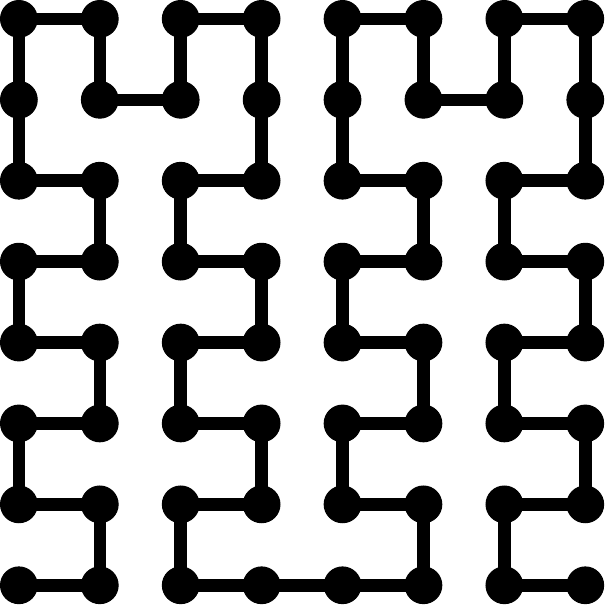}%
}\hfill
\subfloat[\label{10x10Opt}]{%
  \includegraphics[width=0.46\columnwidth]{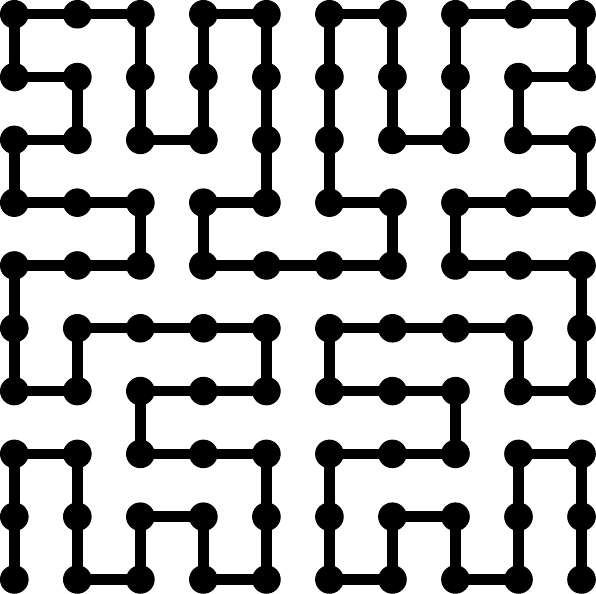}%
}
\caption{%
  \justifying \subref{Moore1} $N=4$: Moore 1, \subref{Moore2} $N=4$: Moore 2, \subref{5x5Generic1} $N=5$: Generic 1, \subref{5x5Generic2} $N=5$: Generic 2, \subref{5x5Generic3} $N=5$: Generic 3, \subref{6x6WorstTiledOpt} $N=6$: Tiled-optimal, \subref{6x6_TiledMixed} $N=6$: Tiled-mixed, \subref{6x6Premature} $N=6$: Premature, \subref{7x7Premature} $N=7$: Premature, \subref{8x8TiledSnake} $N=8$: Tiled-snake, \subref{8x8Premature} $N=8$: Premature, and \subref{10x10Opt} $N=10$: Tiled-optimal.  }
    \label{appendix_paths}
\end{figure}

\section{Computational details} \label{Comp_dets}

All DMRG computations in this work were performed using the julia iTensor library (iTensors v0.9, iTensorsMPS v0.3.18). Additional functions used such as indexing-to-path and tiling/tesselation functions were developed on top of iTensor and are available at \cite{snakeGitHub}.

In order to accurately control the relevant parameters of system size and DMRG path during our analysis, we have thus fixed the initial state, bond dimension vector, cutoff error, and bond dimension increase to be the same in every simulation. We refer to this fixed choice as a protocol. 

The initial state of the MPS was always taken to be the Ne\'el ordered state with the first site in the MPS oriented `down'. As all paths are nearest-neighbour connected by design, every path yields the same check-board pattern of down/up, fixing the initial state. The DMRG cutoff error was fixed to be $10^{-8}$. We have performed 10 sweeps, and the increase vector of the bond dimension is $\{ 2,~4,~6,~8,~10,~12,~16,~32,~64,~128 \}$, increasing the bond dimension accordingly after each sweep. Lastly, we have restricted the computational resources to be the same in every simulation. 

\section{Various Hamilton Paths} \label{App_paths}

\begin{figure}[!b]
\centering
\subfloat[\label{7x7_SE_var}]{%
  \includegraphics[width=\columnwidth,trim={0 1.4cm 0 0 }]{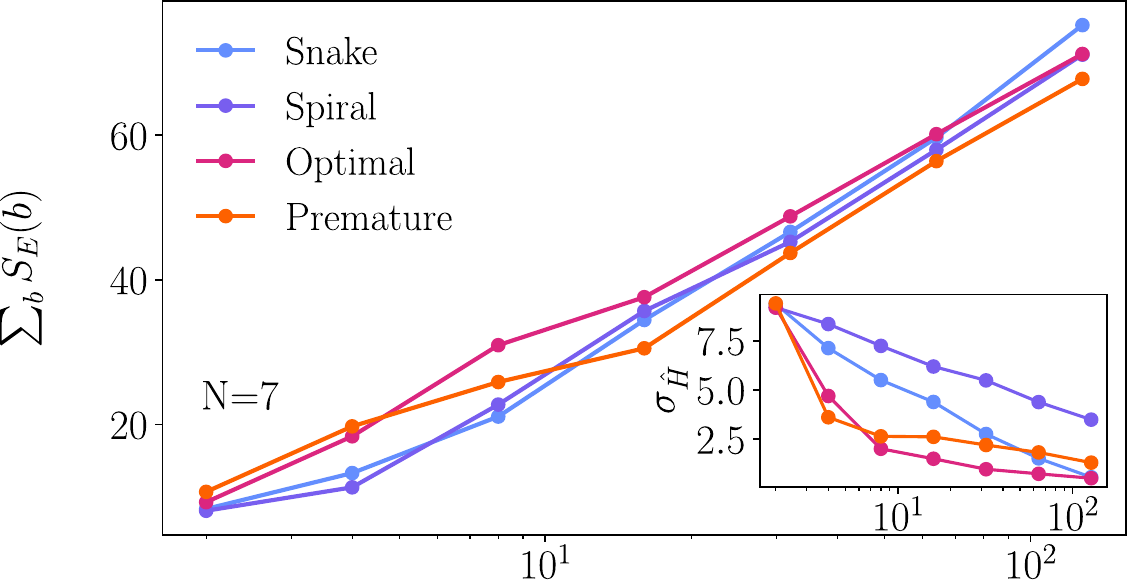}%
}\hfill
\subfloat[\label{8x8_SE_var}]{%
  \includegraphics[width=\columnwidth]{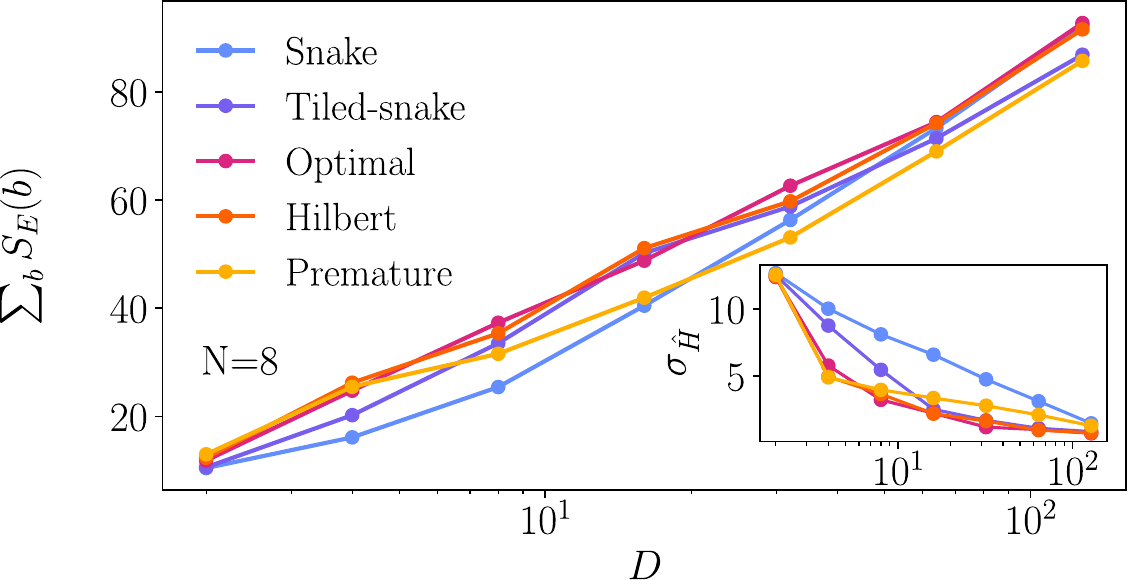}%
}
\caption{%
  \justifying The total bond entanglement entropy of the named paths for the $N=7$ \subref{7x7_SE_var} and \subref{8x8_SE_var} $N=8$ SLHAF. Inset is the variance in the ground state energy for the respective paths.}
  \label{7x7_8x8_SE_var}
\end{figure}

In Fig. \ref{appendix_paths} we show a number of Hamilton paths mentioned in the main text that have been collated here to de-clutter the above discussions. The Hamilton paths associated with the `Moore curve' \cite{Sagan1994SF} on the $N = 4$ grid graph, denoted Moore 1 and Moore 2 are shown in Fig. \ref{Moore1} and Fig. \ref{Moore2} respectively. Three generic paths of the $N=5$ grid graph that show surprisingly good performance are shown in Fig. \ref{5x5Generic1}, Fig. \ref{5x5Generic2}, and Fig. \ref{5x5Generic3}, denoted Generic 1, 2, and 3 respectively. Generic 1  and Generic 2 perform very well for smaller bond dimensions, and Generic 3 performs comparably well to the optimal path, with which it shares structural features (compare to Fig. \ref{Opt5}). The $N=6$ tiled-optimal path shown in Fig. \ref{6x6WorstTiledOpt} is the worst performing path of the four Hilbert tilings of the $N=3$ Peano-Meander motif. The tiled-mixed path in Fig. \ref{6x6_TiledMixed} has lower quadrants that are $N=3$ snake paths and upper quadrants that are $N=3$ Peano-Meander motifs. The premature paths of the $N=6, 7,$ and $8$ lattices are shown in Figs. \ref{6x6Premature}, \ref{7x7Premature}, and \ref{8x8Premature} respectively. The tiled-mixed path shown in Fig. \ref{8x8TiledSnake} is a Hilbert tiling of the $N=4$ snake path. Lastly, the best performing Hilbert tiling of the $N=5$ optimal path is shown in Fig. \ref{10x10Opt}. 

\section{$7\cross7$ and $8\cross8$ entanglement entropy and variance} \label{7x7_8x8_appendix}

Figure \ref{7x7_8x8_SE_var} shows the total bond entanglement entropy and variance in ground state energy of the identified paths for the $N=7$ and $N=8$ square lattices. As for the smaller values of $N$ we again observe the same trends; the premature and optimal paths saturate entanglement faster than the other paths for small $D$ and generally yield a lower variance. For $N=7$ we can see the rate of change of total entanglement entropy of the optimal path beginning to slow, which is likely indicative of the previously observed lower total entropy value for the optimal paths. Such a slowing is not observed for the $N=8$ data, which suggests that larger values of $D$ are required before we can observe the entropy plateau. 

\end{appendix}
\newpage

\bibliography{refs}
\end{document}